\documentclass[twocolumn,trackchanges]{aastex7}

\usepackage{amsmath}
\usepackage{caption}
\usepackage{subcaption}
\usepackage{amsmath}

\def\Cloudy{{\sc Cloudy}}
\def\XRISM{{XRISM}}
\def\Athena{{Athena}}

\begin{document}

\title{\Cloudy{} and the high-resolution microcalorimeter revolution: Optical, UV, and X-ray Spectra of one-electron systems}

\author[0000-0002-4634-5966]{Chamani M. Gunasekera}
\affiliation{Space Telescope Science Institute\\
3700 San Martin Drive, Baltimore, MD 21218}
\email{cgunasekera@stsci.edu}

\author[0000-0001-7490-0739]{Peter A. M. van Hoof}
\affiliation{Royal Observatory of Belgium\\
Ringlaan 3, 1180 Brussels, Belgium}
\email{p.vanhoof@oma.be}

\author[0000-0002-8823-0606]{Marios Chatzikos}
\affiliation{Department of Physics \& Astronomy, University of Kentucky \\ Lexington, KY 40506, USA}
\email{mchatzikos@gmail.com}

\author[0000-0003-4503-6333]{Gary J. Ferland}
\affiliation{Department of Physics \& Astronomy, University of Kentucky \\ Lexington, KY 40506, USA}
\email{gary@g.uky.edu}

\begin{abstract}
With the launch of the \XRISM{} microcalorimeter mission, space-based X-ray observations will achieve a record
spectral resolving power of $R\equiv E/\Delta E\sim$1200. 
With this resolving power, emission features associated with fine-structure energy levels 
of some species will be resolved, sometimes for the first time. 
The plasma code, \Cloudy, was not originally designed for high-resolution X-ray spectroscopy
and throughout its history did not resolve fine-structure components of Lyman lines. 
Here we expand \Cloudy{} to resolve these fine-structure energy levels
and obtain predicted X-ray spectra that match the resolution of new microcalorimeter observations. 
We show how the Lyman lines can be used as column density indicators {in the hot X-ray emitting gas in a cluster of galaxies such as Perseus}, and examine their sensitivity
to external radiation fields and turbulence.
\end{abstract}



\section{Introduction} 

There are a multitude of astrophysical objects that emit X-rays, from galaxy clusters to supernova remnants to X-ray binaries and many more. X-ray astronomy has developed into an extensive field of research, and has made significant strides in understanding the hot and energetic universe. X-ray emission is mainly produced in gas at temperatures from $10^6$ to $10^8$ K, with most detectors working in the $0.1$ to $10$ keV range \citep{2020arXiv200304962X}. 

\Cloudy{} conducts simulations of non-equilibrium plasmas and predicts the entire spectrum including X-ray line intensities. In early \Cloudy{} versions, the emphasis was on producing high-resolution optical, UV and IR spectra, with the X-ray region treated to the precision only required by the then-existing missions \citep{1998PASP..110..761F}. Microcalorimeters allow unprecedented resolution in the X-ray regime, so \Cloudy{} must be improved to match this. This work is part of an effort to improve \Cloudy{} for work in X-ray astronomy (PI: Chatzikos).  Specifically, it aims to resolve the Lyman\footnote{
The Siegbahn notation in X-ray spectroscopy is typically used for inner-shell transitions, while the Lyman series in atomic physics is reserved for hydrogen-like ions. 
In this paper, we will use the latter notation, following the standard practice in X-ray astronomy. So for instance, we will notate the spectroscopic lines arising from 2$p$ $\rightarrow$ 1$s$ for H-like ions as Ly$\alpha$.
} lines into its fine-structure components, which are observable features in microcalorimeter data.

{XRISM observations of Centaurus X-3 Ly$\alpha_1/$Ly$\alpha_2$ ratio for the first time show an unexpected deviation from the expected 2:1 intensity ratio for optically thin emission. Similar Fe XXVI doublet ratios have been observed in the Sun, \citealp{1986PASJ...38..225T}). The numerical and physical advances fully described here were employed by \citet{2025A&A...694L..13G} to reproduce this Ly$\alpha$ ratio in Cen X-3. That Letter also shows that such line ratios can be used as a new column density diagnostic tool for X-ray emitting objects. We present here the microphysical treatment that enables this new column density diagnostic, and provide further diagnostic tools.}

{This paper describes one of several steps in preparing \Cloudy\ for the microcalorimeter revolution in X-ray astronomy.}
\Cloudy{} treats 1 and 2-electron systems with a unified approach along iso-sequences. The two-electron iso-sequence was expanded with and emphasis on optical emission lines  \citet{2012MNRAS.425L..28P, 2013MNRAS.433L..89P}. Subsequently 
\citep{2020ApJ...901...68C, 2020ApJ...901...69C, 2021ApJ...912...26C, 2022ApJ...935...70C} extended the framework to meet the spectral resolution of X-ray microcalorimeter missions. 
For many-electron systems, \Cloudy{} uses atomic databases \citep{2015ApJ...807..118L}, and \citet{2022Astro...1..255G} updated the version of the CHIANTI atomic database used by \Cloudy{}, improving the calculated line wavelengths and intensities. 
{We designed a python script (\url{https://gitlab.nublado.org/cloudy/arrack}) which casts the data from the latest version of Chianti to the format in version 7, stabilizing the Chianti database used by \Cloudy{}.} The upcoming 2025 release of \Cloudy{} (C25) will further update Chianti to 10.1 \citep{2023ApJS..268...52D} using the same method. 

The $np$ subshell of H-like ions is split into two fine-structure levels (with $j = 1/2, 3/2$) by the spin-orbit interaction between an atomic nucleus and the atomic electron \citep{1957qmot.book.....B}. 
In many spectroscopic observations, the $np \rightarrow 1s$ transitions appear as single lines. 
For instance, lines like \ion{H}{1} Ly$\alpha$ are actually doublets although to the best of our knowledge they have never been resolved in astrophysical spectra.
As a result, versions of \Cloudy{} up to now have not resolved one-electron doublets.
As we show below, the doublet spacing increases as nuclear charge ($Z$) increases, and iron-peak elements display two lines that are well separated in microcalorimeter observations.
This work expands \Cloudy{} to predict such transitions, by improving the treatment of one-electron systems in order to match the resolving power of the new X-ray missions.

This paper is organized along the steps we took to resolve the one-electron doublets. In Section~\ref{sec:strategy}, first we outline a strategy that leverages our existing infrastructure, which applies over a broad range of physical conditions and obeys thermodynamic limits under the appropriate conditions. Then in Section~\ref{sec:method} we detail the atomic structure and rates needed to simulate X-ray spectra. Section~\ref{sec:results} discusses several unique radiative transfer issues that arise along the one electron sequence. We show spectra in some simple cases for a simulation of the Hitomi spectrum of the Perseus cluster.

\section{One-electron Doublets in \Cloudy}
\label{sec:strategy}

\subsection{\Cloudy's Previous Strategy}

\Cloudy{} has long treated one and two-electron systems with great care for several reasons. The first is due to the large abundance of hydrogen and helium which, together, make up 99.9\% of the atoms in the universe. 
The second distinction is that the energy-level structures of one and two-electron systems are quite different from the complex energy structure that is found in many-electron systems like, for instance, \ion{O}{2} and \ion{O}{3}. As shown in Figure 3 of \citet{2013RMxAA..49..137F}, the first excited state of one and two electron systems is at roughly 3/4 of the ionization potential, and most of the states are very close 
to the continuum above. These highly excited levels called the Rydberg states in atomic physics, mediate the recombination process, affecting the ionization. Hydrogen and helium, in particular, must be treated with great care because they determine the ionization structure of a cloud (Chapter 2 of \citet{2006agna.book.....O} and their recombination lines are important in determining the composition and ionization of clouds across the universe
(Chapter 5 of \citet{2006agna.book.....O}.

The Rydberg levels pose several interesting problems. An infinite number exists in the low-density limit, although the number of levels is truncated at finite densities due to continuum lowering \citep{2022PASP..134g3001A}. We must sum over all the levels to obtain the total recombination coefficient and predict the ionization correctly. 
The higher levels collisionally couple to the continuum to bring the atom’s ionization into LTE or STE at high particle or photon densities. So, again, many levels must be included. Many strong optical and infrared lines have upper levels in the Rydberg states, so we must determine their level populations with some precision to predict the spectrum. The fundamental problem is to treat a very large number of levels with the available computer hardware.

Our treatment of the Rydberg levels has changed as computers have become faster. Initially, we used several pseudo-states to represent the closely-spaced Rydberg levels at high principal quantum numbers \citep{1987PhDT.........7C, 1997ApJ...479..363F}. The pseudo-states allowed the atom to go to LTE and STE limits when the particle or photon densities were sufficiently high \citep{1988ApJ...332..141F, 1989ApJ...347..656F}. A disadvantage to this approach was that the pseudo-states affected the accuracy of the H and He recombination-line intensities. 
This approach reproduced classical Case B  \citep{2006agna.book.....O} intensities of H and He recombination lines to better than five percent.

Recombination lines must be predicted to high precision  for certain applications, such as the primordial helium abundance of the universe, \citep{2010IAUS..268..163F} or for denser environments, where collisional and radiative transfer effects may be important \citep {1999ASPC..162..147F}. Classical case B productions do not describe such clouds, so detailed radiative transfer must be done simultaneously with the solution of the emission and ionization. 

As computers became faster, it became possible to remove the pseudo-states and replace them with models of higher-$n$ shells. This advance was described in a series of papers that focused on measurements of the primordial helium abundance \citep{ 2005ApJ...628..541B, 2005ApJ...622L..73P, 2007ApJ...657..327P, 2009MNRAS.393L..36P, 2012MNRAS.425L..28P}. \citet{2013RMxAA..49..137F} is the culmination of this development. Figure 1 of that paper shows our model for one-electron systems. We use $nl$-resolved states for low principal quantum numbers. ``Collapsed states'', which are not $l$-resolved, were used for high $n$ where $l$-changing collisions should bring the $nl$ populations into $g = 2l+1$ statistical equilibrium.

Pseudo-states are no longer used to describe the recombination and line-producing physics in our current approach. Beginning with C13, we use a finite number of collapsed and resolved levels,  with a small amount of ``top off'' recombination coefficient being added to the highest level to reproduce the total recombination to all levels.
This approach had a problem near photoionization edges such as the Lyman jump. This is shown in figures 7 and 8 of \citet{2017RMxAA..53..385F}. 
In nature, the very high-$n$ Lyman lines merge onto the Lyman continuum above the bound levels, and no discontinuous Lyman jump is present \citep[Section 3.1.4]{1969mpap.book.....B, 2017RMxAA..53..385F}. A finite model produces gaps in the spectrum just longward of the ionization edge where unmodelled high-$n$ lines should add opacity. The continuum can ``leak'' through the cloud, as shown in Figure 8 of that paper. We dealt with this by adding many “extra” Lyman lines. These added opacity to the cloud but, in the original treatment, did not produce emission. The extra Lyman lines were sources of absorption opacity, so their upper-level population need not be known and they were not included in our level-population solver.

As outlined above, the original treatment of one-electron systems focused on light species such as \ion{H}{1} or \ion{He}{2}. We treat one and two electrons systems with a unified model that extends to the heaviest element treated, currently zinc ($Z=30$). This is coded in such a way that it could be extended to very heavy elements if sufficient atomic data were available.

The following paragraph provides a brief overview, while subsequent sections will dive into greater detail. As described in the Introduction, high-resolution X-ray spectroscopy is becoming commonplace. The Lyman lines in one-electron systems are doublets. The discussion in the next section shows that the doublet separation is small for light elements, such as hydrogen and helium, and would not be resolvable for typical astrophysical kinetic temperatures ($\sim 10^4$\,K). Previous developments had resolved $nl$ but not the $j$-levels that introduce the doublet splitting. The doublet separation increases with the nuclear charge $Z$. Microcalorimeter X-ray missions will resolve the Ly$\alpha_{1,2}$ doublets for elements heavier than calcium \citep{2025A&A...694L..13G}. The doublet separation depends on $n$. It is largest for the 2-1 transition and decreases as $n$ increases. Future X-ray missions will resolve lower-$n$ transitions of higher-$Z$ species. \Cloudy{} has long treated up to zinc. However, this treatment can be extended to any $Z$ and any principal quantum number $n$ that the user specifies.

This paper will further develop the extra Lyman lines described in \citet{2017RMxAA..53..385F} to predict doublet emission. 
We use the existing level and ionization population solvers, which are $nl$ but not $j$ resolved, to determine the populations of upper ($^2$P$_j$) levels and include the emission that results. We show synthetic spectra of Hitomi's observations of the Perseus cluster.

\subsection{Overview of the New Strategy}

We adopted a strategy to resolve the Lyman doublets within the pre-existing framework. In Section~\ref{sec:method}, we describe each step in greater detail.

Extensive tests show that the existing one-electron populations solvers go to all thermodynamic limits. This includes LTE at high densities, STE when exposed to a true blackbody, the highly ionized Compton limit, and the fully molecular limit where most H is in the form of H$_2$. We refer to this existing solver as the full collisional-radiative model (CRM) solver. The goal is to retrofit the doublets into this scheme.

To match the resolving power of the new microcalorimeter X-ray missions (the \textit{X-ray imaging and spectroscopy mission}, \XRISM, and the \textit{Advanced telescope for high-energy astrophysics}, \Athena), we need to self-consistently resolve the single $np \rightarrow 1s$ lines in \Cloudy{} into the fine-structure $j$-resolved doublets within \Cloudy's existing framework. Below we will use the following notation for the Lyman doublets. The $2p_{1/2} \rightarrow 1s_{1/2}$ line will be written as Ly$\alpha_2$ and the $2p_{3/2} \rightarrow 1s_{1/2}$ transitions as Ly$\alpha_1$. Similar notations will be used for the higher Lyman lines.

We begin with the well-known theory of radiative transfer. A beam of radiation with energy $h\nu$ propagating through a medium has an intensity $I_\nu$ that evolves according to the well-known equation of radiative transfer:
\begin{equation}
    dI_\nu = -I_\nu \kappa_\nu ds + j_\nu ds,
\end{equation}
where the beam has traversed a path $s \rightarrow s+ds$, j$_\nu$ and $\kappa_\nu$ are the emission and absorption coefficients at frequency $\nu$ respectively. Atoms, ions, and molecules can absorb and emit radiation of frequency $\nu$, contributing to the emission and absorption coefficients as follows:
\begin{equation}
    j_\nu = (1/4\pi) n_u A_{ul}h\nu \phi_\nu,
    \label{eq:emission_coeff}
\end{equation}
\begin{equation}
    \kappa_\nu = n_l \sigma_{lu}(\nu) - n_u\sigma_{ul}(\nu), \\ \sigma_{lu}\propto\frac{g_u}{g_l}\frac{1}{\nu_{lu}^2}A_{ul} \phi_\nu
    \label{eq:absorption_coeff}
\end{equation}
where $n_u$ is the population density of the level $u$, $A_{ul}$ is the transition probability, $u$ and $l$ are the upper and lower energy levels of the transition, $\sigma_{lu}$ is the absorption cross-section, and $\phi_\nu$ is the normalized line profile. 

\begin{figure}
    \centering
    \includegraphics[width=\columnwidth]{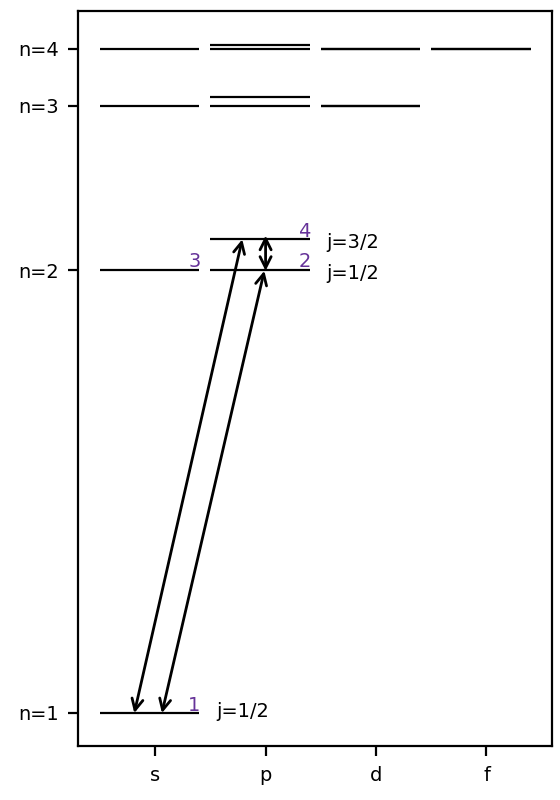}
    \caption{Energy-level diagram showing the fine-structure levels of angular momentum quantum number $l=1$. The level ordering is consistent with energy levels of lighter H-like species. 
    We denote energy levels $1s$, $2p$ ($^2P_{1/2}$), $2s$, and $2p$ ($^2P_{3/2}$) shown here simply as levels 1,2,3 and 4 respectively in the text.}
    \label{fig:E_level_diagram}
\end{figure}

Figure~\ref{fig:E_level_diagram} shows an energy level diagram of a one-electron atom, with the key electron transitions that need to be considered. In concurrence with this figure, for the remainder of this manuscript, we will denote energy levels $1s$, $2p$ ($^2P_{1/2}$), $2s$, and $2p$ ($^2P_{3/2}$) simply as levels 1,2,3 and 4 respectively. The above theory reveals that first, we need only to calculate the population densities $n_{2}, n_{4}$ and the frequencies $\nu_{12}, \nu_{14}$ of the $j$-resolved levels/transitions. Then \Cloudy's existing solvers, which compute the total 1$s$, 2$p$, and 2$s$ populations, would self-consistently produce the correct line intensities. The equation of detailed balance in steady state shows us how $n_{2}$, $n_{4}$ relates to the population density $n_{2p}$ already in \Cloudy,
\begin{equation}
n_{2p} A_{2p,1} \beta_{net} = n_{4} A_{4,1} \beta_{4} + n_{2} A_{2,1} \beta_{2},
\label{eq:detailed_balance}
\end{equation}
where $\beta$ is the escape probability accounting for radiative trapping effects. In Section~\ref{sec:pop_den}, we detail our population density determinations for the $j$-resolved states. 

The line frequencies $\nu_{12}, \nu_{14}$ are set by the fine-structure resolved level energies. Spin-orbit coupling combined with relativistic and quantum electro-dynamic effects lift the degeneracy in the $nl$ energy levels, splitting $^2P$ levels into two \citep{1957qmot.book.....B}. Section~\ref{sec:energy} discusses the complete energy calculations and the physics behind fine-structure levels. 

Next, we need to account for line overlap in our line profile function. Equations~\ref{eq:emission_coeff}, \ref{eq:absorption_coeff} and the escape probability $\beta$ make use of a single line profile for the $2p \rightarrow 1s$ transitions.  \citet{2005ApJ...624..794S} developed a multi-grid approach where the gas opacity is treated with two energy meshes. Much of the continuum radiative transfer is done with our ``coarse continuum'', which has a resolving power of order $10^3$. The fine opacity mesh has far higher resolution and resolves the line profiles, allowing for an automatic treatment of line overlap. This was necessary due to the density of electronic H$_2$ lines at photodissociation energies.

Section~\ref{sec:escape_prob} discusses overlapping lines in greater detail
for the H Ly$\alpha$ line and using the sum of the optical depths of the $j$-resolved lines for Ly$\beta$ and higher lines. The calculation of the $j$-resolved escape and destruction probabilities and their relation to the non-$j$ resolved lines, is also presented here.

Since resolving $np \rightarrow 1s$ for all Z and $n$ is not needed in every case, we use the astrophysical context and the energy spectral resolution of \XRISM{} and \Athena{} to select the reported $Z,n$. We use a default spectral resolution to resolve the fine-structure doublets and introduce a new user command to change it. This is discussed in Section~\ref{sec:resolution}.

We now know the individual opacities for the $j$-resolved lines and need to get this information into \Cloudy's main $nl$ solver self-consistently. For each microphysical process that \Cloudy{} simulates (such as line trapping,
continuum fluorescence, destruction by background opacities), we redirect the main solver to use the $j$-resolved physical quantities (populations, energies, opacities) for each $Z$ and $n$. 

The energy of the 2$s$ ($^2$S$_{1/2}$) level is very close to the 2$p$ ($^2$P$_{1/2}$) level. Most 2$s$ decays produce two-photon emission but a magnetic dipole single photon transition $2s_{1/2} \rightarrow 1s_{1/2}$ (hereafter the M1 line) is possible. Appendix \ref{sec:M1} compares the two- and one-photon rates. The M1 line has an energy that is close to the $j$-resolved transition $2p_{1/2} \rightarrow 1s_{1/2}$. Thus, when resolving the fine-structure doublets as discussed above, there is some ambiguity with the M1 line resulting in M1 line contributions to the intensity of the $j$-resolved doublet, the classical Ly$\alpha$ transition. We disambiguate these lines by giving the M1 lines in \Cloudy{} a new line label. This is further discussed in Section~\ref{sec:M1}.

The 2$p$ fine-structure levels of two-electron systems are singlets and triplets. Our model two-electron atom resolves these into $nl$-resolved states for all levels. The $2 ^3$P level is split into its three $j$ levels since these are important X-ray diagnostics for higher-$Z$ elements \citep{2007ApJ...664..586P}. As such, the treatment of the He-like ``extra'' Lyman lines remains unchanged from the previous versions of the code. The fine-structure splitting of subordinate lines is far smaller than for the resonance lines so they will either be in the unobservable XUV or unresolvable in astrophysical applications.

\section{Atomic Structure and Rates}
\label{sec:method}

The CRM solver adopts $nl$-resolved energy levels. To fit the $nlj$-resolved fine structure calculations into the existing $nl$-resolved infrastructure, we take the following steps. First, we will duplicate the pre-existing one-electron ``extra'' Lyman line structure to be used for the two $j$-resolved arrays $np_j$, for $j=1/2$ and $j=3/2$. Using the CRM $nl$ level populations, we  populate the upper levels of the $nlj$ array, with the appropriate population densities (further discussed in Section~\ref{sec:pop_den}). The CRM solver evaluates the total doublet emission lines from $np$ to $1s$. We recover the $nlj$-resolved line optical depths from the known relationships among the Einstein rate coefficients.

The astrophysical context and instrumental limits determines which H-like ions and levels should be fine-structure resolved. It is also necessary to determine whether j-changing collisions are important in the density domain utilized by \Cloudy. The following subsections discuss our analysis, and determinations of each of these values.

\subsection{Energy Calculations}
\label{sec:energy}

The spin-orbit interaction between an atom's electron and its nucleus is a perturbation that lifts the degeneracy of energy levels by splitting states with different quantum numbers $l$ and/or $j$. In one-electron systems, the $^2$P state ($l=1$) is resolved into a doublet with $j=1/2$ and $j=3/2$.

We use the following approximation for the binding energy $E_{np}$ of an $np$ electron in a hydrogen-like atom. Here we retain the lowest-order correction terms, which will give sufficient accuracy for our needs.
\begin{equation}
    E_{np} = {E_n^0} + E^{FS}_{nj} + E^{\rm LS}_{n,l=1,j} + E_{nj}^{M}-E_{ion},
    \label{eq:totalE}
\end{equation}
$E_{ion}$ is the ionizing potential of the one-electron ion (which by definition is for $n=1$) already given within \Cloudy{} (obtained from \citealp{2015JPCRD..44c3103Y}).
Here, $E_n^0 + E^{FS}_{nj}$ is the fine-structure energy that resolves the levels with different j given in Eq.~\ref{eq:dirac}. 
$E_{n,l>0,j}^{\rm LS}$ is the Lamb Shift correction which resolves the levels with different l, and $E_{nj}^{M}$ is the nuclear mass recoil correction, given in Eq.~\ref{eq:E_LS} and \ref{eq:E_M} respectively. A discussion of the full calculations of all these terms used in \Cloudy{} are provided in Appendix~\ref{app:energy}.

\subsection{Energy Resolution \& Accuracy}
\label{sec:resolution}

\begin{figure}
    \includegraphics[width=1.\columnwidth]{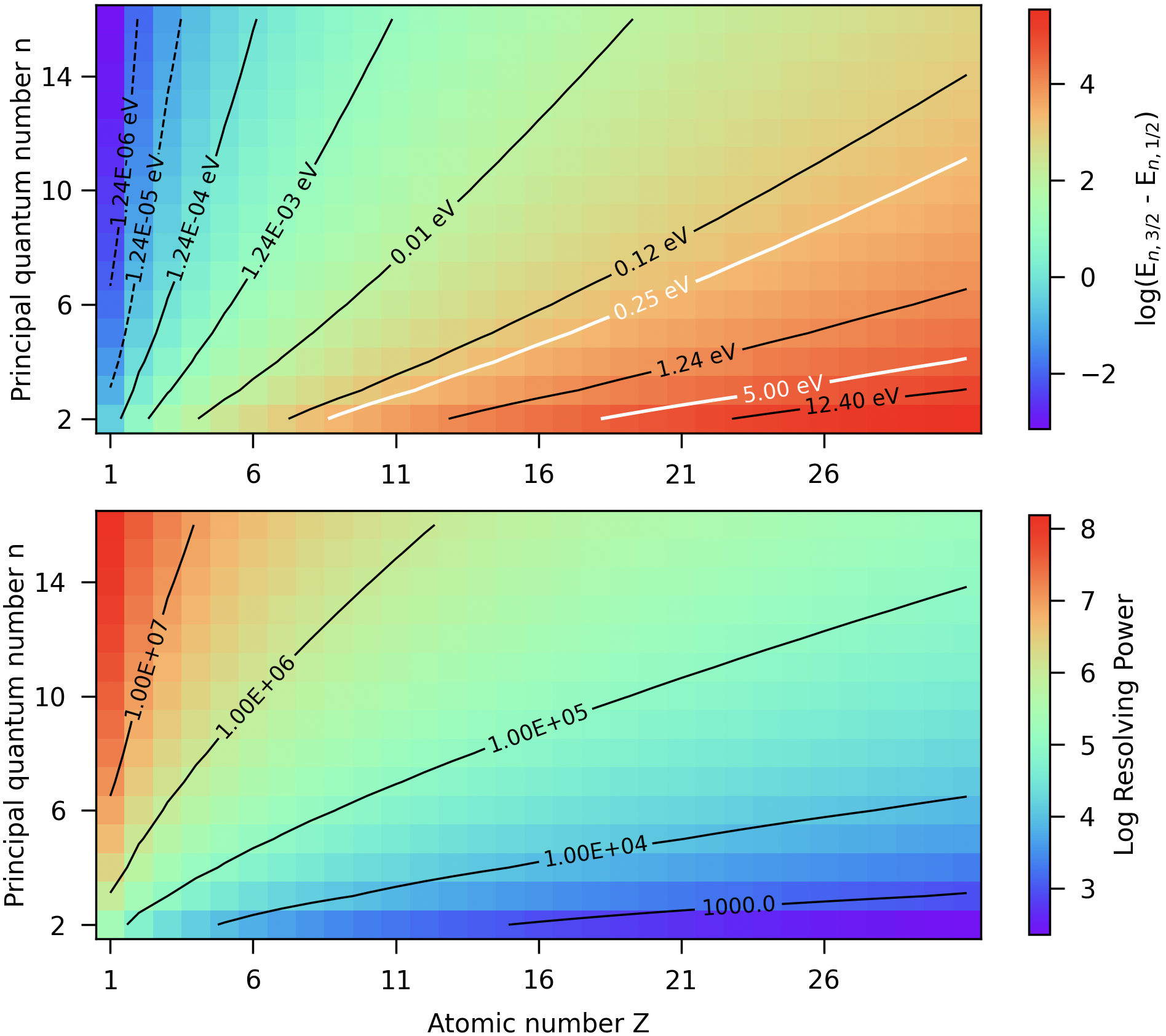}
    \caption{\textit{Top:} Contour plot of energy separation between the $np$ ($^2$P$_{1/2}$) and $np$ ($^2$P$_{3/2}$) levels. The white plot lines indicate the spectral resolution of the current microcalorimeter mission \emph{XRISM} (R$\sim$5 eV) and future mission \Athena{} (R$\sim$2.5 eV). \textit{Bottom:} Contour plot showing the resolving power required to distinguish between the $np_{1/2}$ -- $1s_{1/2}$ and $np_{3/2}$ -- $1s_{1/2}$ transition energies, where E$_{1s_{1/2}}=0$. For both plots the y-axis is the range of principal quantum numbers used in \textsc{cloudy} calculations, and the x-axis is the range of elements with atomic number Z used.}
    \label{fig:energyResPlot}
\end{figure}
Figure~\ref{fig:energyResPlot} shows a contour map of the spectral energy resolution (top panel), and the required resolving power ($R$) (bottom panel) for fine-structure splitting in the $^2$P shell of nuclear charge ranging from H to Zn ($Z$=30) and $n$ ranging from 1 to 16. The energy difference in the split fine-structure levels become rapidly smaller at higher n levels, but rapidly larger with heavier atomic nuclei. 

\XRISM{} has an energy resolution of 5 eV for the energy range 0.3-12 keV, while \Athena{} is expected to have a resolution of 2.5eV \citep{2020arXiv200304962X}. So to meet the upcoming instrumental requirements, according to figure~\ref{fig:energyResPlot}, we need to resolve the $^2$P shells into doublets for atoms heavier than phosphorus ($Z$=15) and for $n < 5$. However, we use Equations~\ref{eq:dirac}-\ref{eq:N} to calculate the energies for all $Z$ and $n$, and report only those lines resolvable by a given spectral resolution. The reported $j$--resolved lines are determined within the code, using a test comparing the energy difference between the two fine-structure levels to the desired spectral resolution. By default we implement a spectral resolution that is a factor of 10 better than \Athena's predicted resolution (i.e., we use a resolution of $2.5$ eV$/10$ = 0.25 eV), for the Lyman lines. We also introduce a new command allowing users to alter this default resolution (See Hazy 1, Section 12.4.12 Database H-like Lyman extra resolution, of the C25 release).

\begin{figure}
    \centering
    \includegraphics[width=\columnwidth]{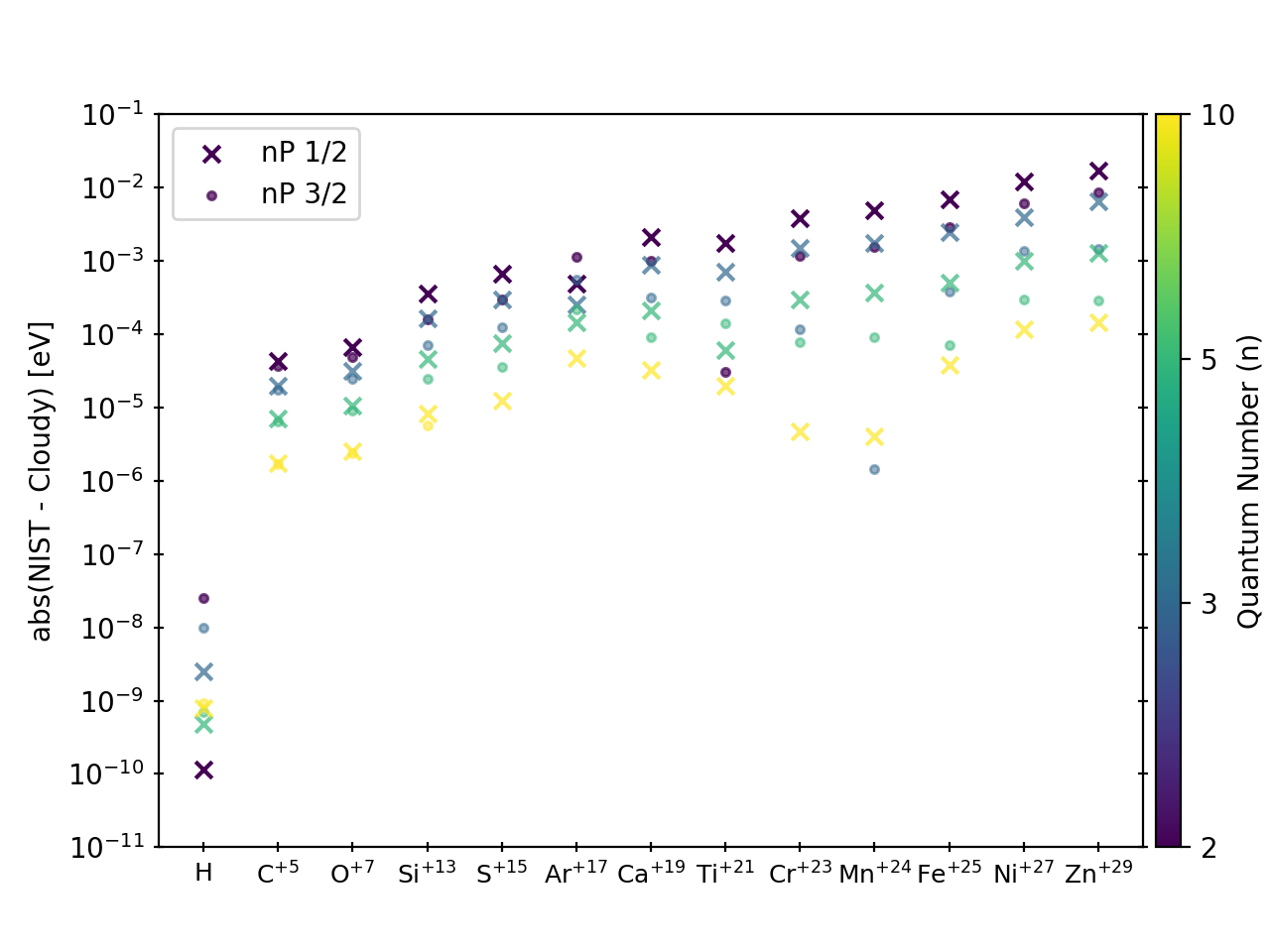}
    \caption{Energy scale accuracy of the updated $np$ ($^2$P$_{1/2}$) and $np$ ($^2$P$_{3/2}$) levels for H-like species in \Cloudy{}, using NIST as the authority. The expected accuracy for \emph{XRISM} is 0.5 eV, which is well above those for any of our new H-like $np_j$ energy calculations. The dots indicate the accuracy for $j=1/2$, and crosses indicate those for $j=3/2$. The colors going from purple to green indicate increasing principal quantum number n.}
    \label{fig:dE}
\end{figure}
Additionally, \XRISM{} has an energy scale accuracy of 0.5 eV \citep{2020arXiv200304962X}. The spectral energy resolution reflects the instrument's ability to differentiate between two closely spaced energies, while the accuracy resembles the instruments ability to detect the energy of a photon as close to its true value as possible. Figure~\ref{fig:dE} provides the difference between our total level energies and those in NIST, in units of eV for various one-electron atoms. We find that our largest energy error is approximately 0.01 eV, which is more than a factor of 10 better than \XRISM's energy accuracy.

This treatment fulfils the energy accuracy and resolution requirements for the up-coming X-ray missions and beyond. We also, prevent fine-structure splitting of the $^2$P level in hydrogen and helium, due to the instrumental limitations as seen from Figure~\ref{fig:energyResPlot}. 

\subsection{Transition Rate Coefficients}
\label{sec:transition_prob}

\Cloudy{} uses non-relativistic calculations to determine radiative transition rate coefficients (a.k.a Einstein A values)
\citep{2006sham.book.....D}. 
These computations are not $j$-resolved, since the correction factors evaluate to unity in our simple case. 
We leave them as they are for now. Future work will involve including relativistic corrections to the \Cloudy{} computed transition probabilities. The current Einstein A's in \Cloudy{} differ from the NIST values which include relativistic corrections, 
by at most $\sim$2 per cent \citep{2004JPCRD..33.1059J}, this accuracy is sufficient for our current instrumental needs.

\subsection{Populations and Intensity Ratios}
\label{sec:pop_den}

The intensities of the emission line are determined by the population density of the corresponding upper level $u$ of the transition ($n_u$) and the rate of spontaneous emission (A$_{ul}$) of the transition $u \rightarrow l$. So, to determine the n$p$ fine-structure line intensities, we need to first determine the population densities of the n$p_{j=1/2}$ and n$p_{j=3/2}$ levels. Here we discuss the prescriptions for $n=2$ levels for simplicity, however, the same framework is extended for levels $n>2$. 

The presence of fine-structure transitions is determined by the density of the ionized gas ($n_{\rm gas}$). 
In steady state, $\frac{dn_{\rm u}}{dt}=0$ results in two possible limits based on how the ionized gas density compares with the critical density ($n_{\rm crit}$),
\begin{equation}
    n_{\rm crit} = A_{\rm u'l'}/q_{\rm lu},
\end{equation}
where, $q_{lu}$ is the rate of collisional excitations from the lower level $l$ to the upper level $u$.

In the low-density limit ($n_{\rm gas}<n_{\rm crit}$), radiative emission is faster than the rate of collisions. Hereby, we will call this the radiative limit.
Here, the density of our gas is sufficiently low enough to ignore collisional $j-$changing transitions, so the $j$-changing transitions can be neglected. 
Here the population ratio is related to the rates at
which the $nlj$ levels are populated.

In the high-density limit collisions are much faster than spontaneous emissions ($n_{\rm gas}>n_{\rm crit}$). We will refer to this limit as the collisional limit. 
Here, $n_2/n_4$ becomes equal to the ratio of the statistical weights of the corresponding levels.

\begin{figure*}
    \centering
    \begin{subfigure}{0.495\textwidth}
        \centering
        \includegraphics[width=\textwidth]{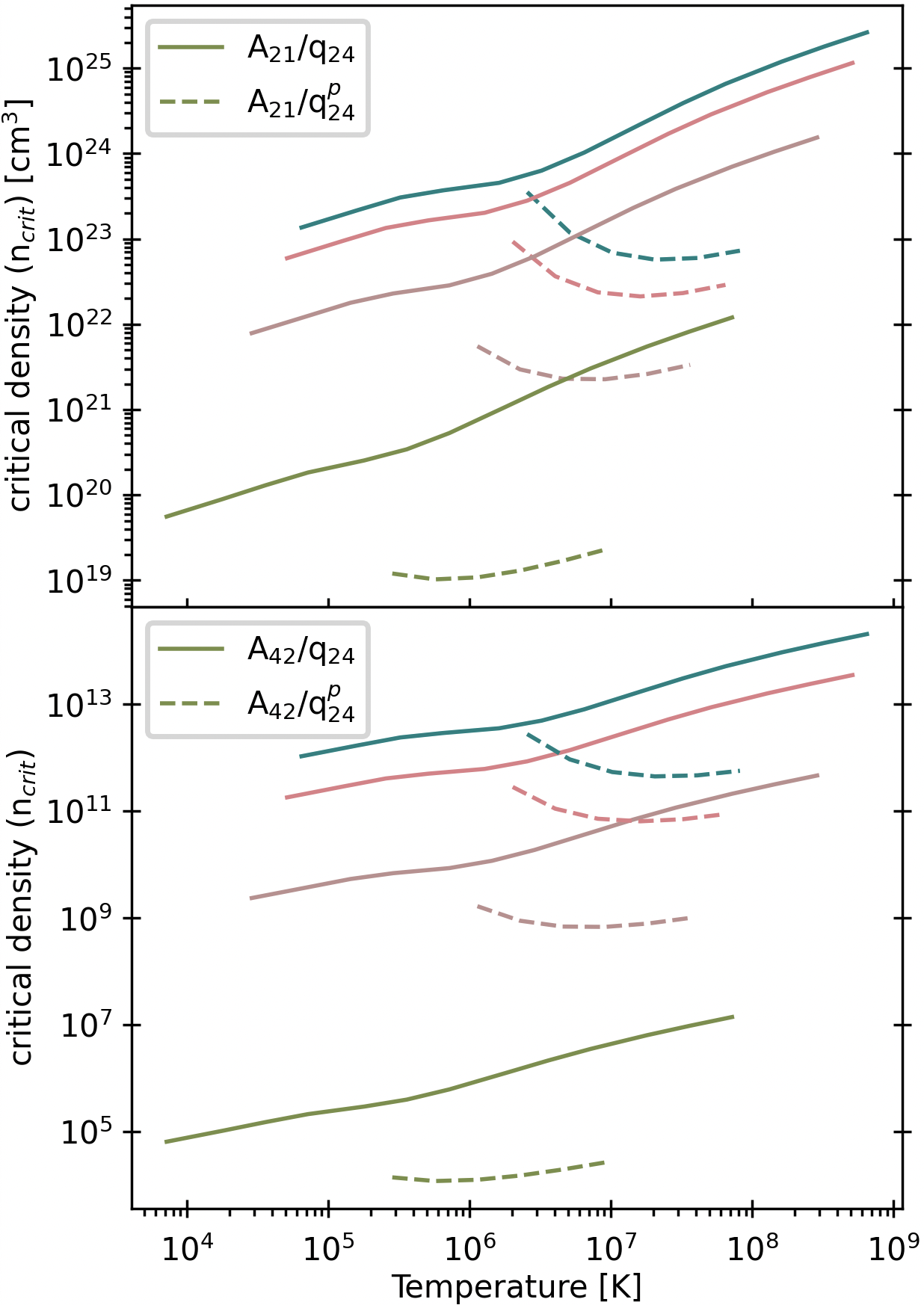}
        \caption{}
        \label{fig:ncrit_a}
    \end{subfigure}
    \hfill
    \begin{subfigure}{0.495\textwidth}
        \centering
        \includegraphics[width=\textwidth]{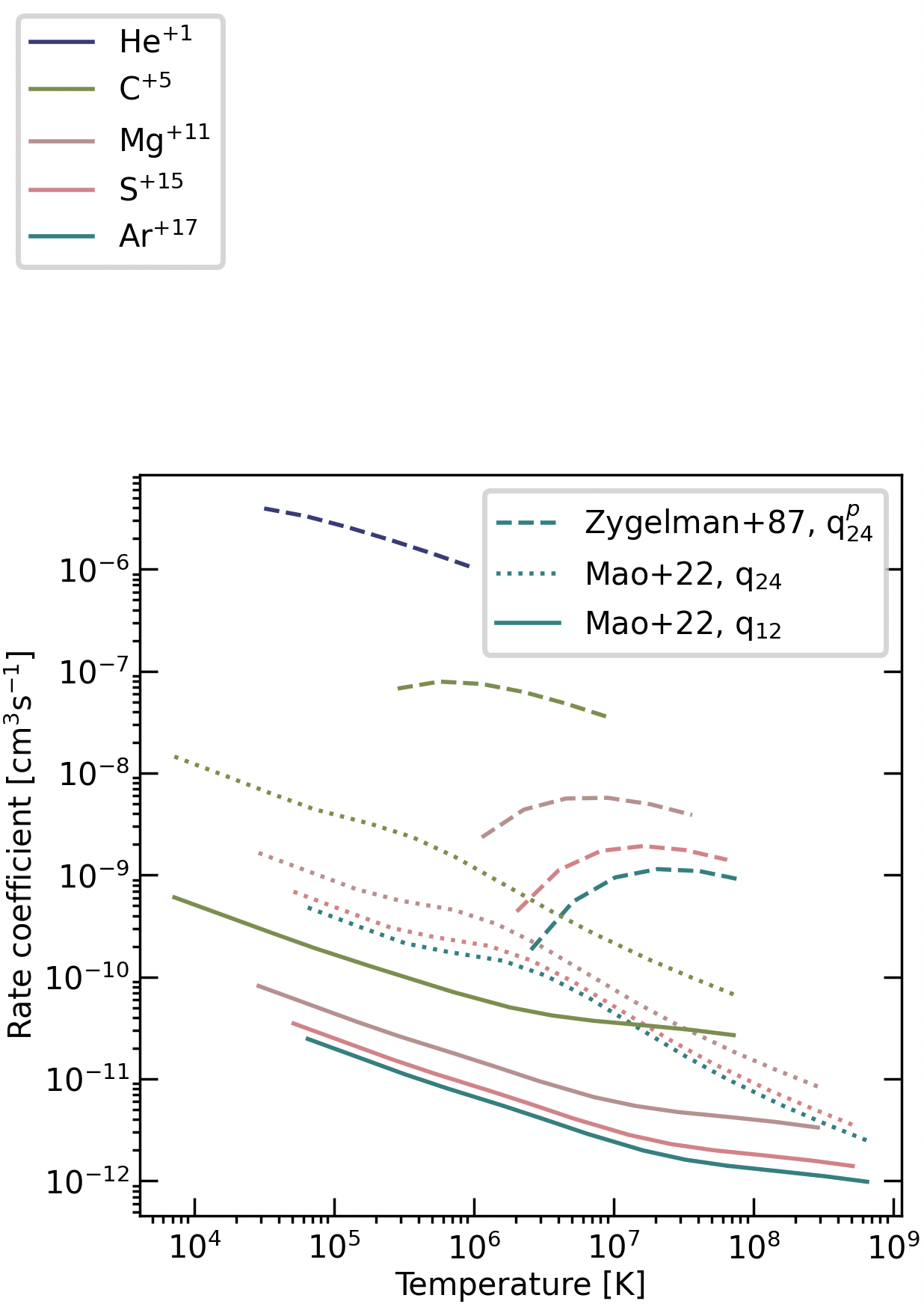}
        \caption{}
        \label{fig:ncrit_b}
    \end{subfigure}
    \caption{(a) $n_{\rm crit}$ for proton- and electron-impact collisions for four one-electron ions. \textit{Top}: ratio of the rate of j-changing collisional excitations ($q_{24}^{i}$) to the spontaneous emission coefficient for the $2\rightarrow1$ transitions. \textit{Bottom}: ratio of the rate of j-changing collisional excitations to the spontaneous emission coefficient for the j-changing ($4\rightarrow2$) transitions. (b) rate coefficients used to compute $n_{\rm crit}$ as a function of temperature for the one-electron ions with proton-impact collisional data.}
\end{figure*}
Figure~\ref{fig:ncrit_a} shows $n_{\rm crit}$ for proton- and electron-impact collisions for four one-electron ions (C\,{\sc vi}, Mg\,{\sc xii}, S\,{\sc xvi}, Ar\,{\sc xviii}), and Figure~\ref{fig:ncrit_b} shows the rate coefficients used to compute $n_{\rm crit}$. We provide electron rates for the $2\rightarrow1$ transition ($q_{12}$) for a more complete comparison.
The lack of proton collision rate coefficients is a pressing atomic data need for the next generation of X-ray observatories. For nearly degenerate energy levels, slow collisions are more effective than fast ones \citep[Section 13.1]{1998clel.book.....J}.
So, electron collisions are less effective at depopulating the $j=3/2$ level by j-changing transitions because of their greater speed in a thermal gas \citep{2025Atoms..13...44D}.
Extensive electron fine structure collision data are given in \citet{2022ApJS..263...35M}.
These were used to calculate the electron rates with the following \citep[Eq. 4-4,4-11]{1978ppim.book.....S}
\begin{equation}
    q_{\rm ul}
               =\frac{8.63\times10^{-6}\Omega(l,u)}{g_uT^{1/2}} 
               = q_{lu}\frac{g_l}{g_u}\exp \left( \frac{E_{lu}}{k_{B}T} \right) 
               {\rm [cm^3s^{-1}]}
\end{equation}
where $g_i$ is the statistical weight of level $i$, $T$ is the temperature of the ionized gas, $\Omega(l,u)$ is the collision rate, $E_{lu}\equiv E_u-E_l$ is the transition energy, and $k_B$ is the Boltzmann constant.
However, we expect proton rates to be more important at lower temperatures as well \citep{1968ApJ...152..701B, 1987PhRvA..35.4085Z}. 
A general theory for ion-ion collisions is given by \citet{1964MNRAS.127..165P, 1988PhR...162....1W}, while \citet{2003JPSJ...72.1073I} give cross sections of proton-impact excitation between the $n=2$ fine-structure levels of hydrogenic ions. 

\Cloudy{} is designed to operate over a very broad range of densities, going from the low-density limit up to LTE, $n \ge 10^{20}$ cm$^{-3}$ depending on the ion, Figure~\ref{fig:ncrit_a} confirms we may reasonably assume the radiative limit and that j-changing collisions can be ignored for densities below $10^{20}$ cm$^{-3}$. The critical densities for all hydrogenic species heavier than C\,{\sc vi} shown in Figure~\ref{fig:ncrit_a} is above $10^{20}$ cm$^{-3}$. 
The gas densities of typical H\,{\sc ii} regions are $\sim10^4$ cm$^{-3}$ which is well below the above-mentioned values of n$_{\rm crit}$. 
This introduces a new uncertainty. In this low-density limit, the $j$-resolved $^2$P level populations are determined by the rates that the various $j$-resolved levels are populated by. That, in turn, would require fine-structure resolved rate coefficients for recombination, collision, and radiative transitions that populate the $j$ levels. Those data are yet to be implemented into \Cloudy{}.

We assume that the $j$-resolved population densities to scale according to the ratio of statistical weights, compared to the population density calculated for the unresolved lines,
\begin{equation}
    n_{npj} = 
    \begin{cases}
    n_{n} \left(\frac{g_{np}}{2n^2} \frac{g_{npj}}{g_{np1/2}+g_{np3/2}}\right), & {\rm collapsed\ states}\\
    n_{np} \left(\frac{g_{npj}}{g_{np1/2}+g_{np3/2}}\right), & {\rm resolved\ states}
    \end{cases}
\end{equation}
where $g_{np} = 2(2l+1) = 6$ is the statistical weight for the $l$-resolved levels with $l=1$, $g_{npj} = 2j+1$ are the statistical weights of the $j$-resolved levels and $j=\frac{1}{2},\frac{3}{2}$. 

\subsection{Radiative Trapping and Line Overlapping}
\label{sec:escape_prob}

Nearly all lines in \Cloudy{} are instances of a C++ object, allowing these lines to be treated with a common code base. Line transfer is done using the escape probability formalism \citep{1984mrt..book.....K, 1992ASSL..170.....E}. This unified code base allows most lines to include radiative trapping and fluorescence, and line thermalization. The code uses various theories to calculate the escape probability, depending on the classification of the line (see, e.g., Section 3.4 in \citealp{Rutten03}). For hydrogen Ly$\alpha$, the code uses fits to the results presented in \citet[hereafter HK80]{1980ApJ...236..609H}, which takes line destruction by background opacities into account.

In general, radiative damping can be important for permitted lines in high-$Z$ elements. The Voigt profile function accounts for this broadening, in addition to broadening of the upper and lower levels of the transition and thermal broadening. \Cloudy{} calculates the Voigt profile using the theory described in \cite{1999JQSRT..62...29W} and \cite{1938ApJ....88..508H}. This routine is very accurate.

The HK80 theory implicitly assumes the line to be single, i.e., it does not consider the case where lines overlap, as occurs for the two fine-structure components of the Ly$\alpha$ line. In the following, we will show how we modified the theory to deal with this problem.

The starting point of the theory presented in HK80 is the $\beta_{HK}$ parameter, the ratio of background continuum ($k_c$) to line
opacities, which is defined as (Eq.~2.7 of HK80):
\begin{equation}
    \beta_{HK} \equiv k_{\rm c}/k_{\rm L},
\end{equation}
\Cloudy{} is designed to work over a very broad range of physical conditions, including cases where the continuous opacity is large. We work with a modified version, the ratio to total opacity,
\begin{equation}
    \beta_{HK}^{'} = k_{\rm c}/\left(k_{\rm c} + k_{\rm L} \right) .
\end{equation}
For most clouds, the two versions of $\beta_{HK}$ are nearly equal. $\beta_{HK}^{'}$ is the probability that a photon will be absorbed and destroyed by background opacity rather than by line scattering.

The frequency-dependent line opacity $k_{\rm L}$ is defined as (Eq.~2.3 of HK80):
\begin{equation}
    k_{\rm L} = \frac{N_l B_{lu} h \nu_0}{4\pi \Delta},
    \label{kldef}
\end{equation}
where $N_l$ is the population of the lower-level, $B_{lu}$ is the einstein B coeffient of the transition $u\rightarrow l$, $h$ is the Planck constant, $\nu_0$ is the central
frequency of the line, and $\Delta$ is the Doppler width of the line.
This can be simplified to Eq.~8 of \citet{2023RNAAS...7..246G}:
\begin{equation}
    k_{\rm L} = N_l \kappa_{\rm L} \sqrt{\pi} / \Delta_{\rm v},  \\ {\rm where} \\  \Delta=\frac{\nu_0}{c}\Delta_{\rm v}.
    \label{kldef3}
\end{equation}

\begin{figure}
    \centering
    \includegraphics[width=\columnwidth]{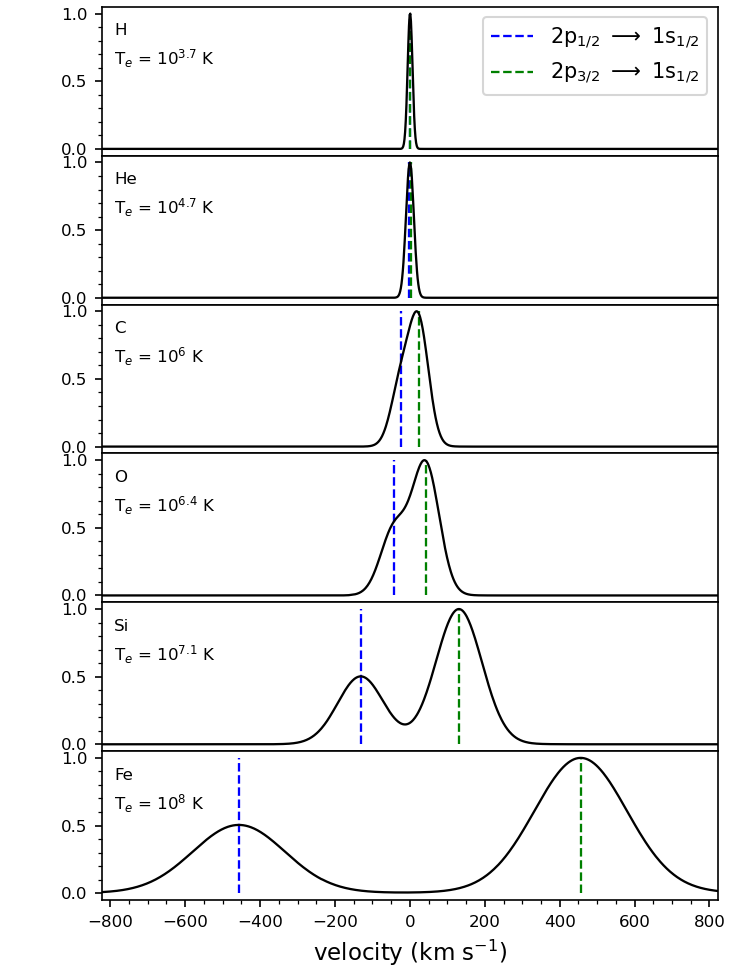}
    \caption{Normalized line opacities as a function of line-of-sight velocity for several one-electron 2$p$ fine-structure doublets, at a spectral resolution of 0.25 eV, showing the doublet splitting increasing with increasing $Z$. The blue and green dashed lines indicate the position of the $j=1/2$ and $j=3/2$ lines, respectively. The temperature of the gas is indicated in the top left corner of each panel and corresponds
    to the temperature where the ion's abundance peaks in 
    collisional equilibrium. Line profiles tend to become
    sharper as the nuclear mass increases but broader as the
    temperature increases.}
    \label{fig:LineOverlap}
\end{figure}

\Cloudy{} uses a multi-grid approach to fully resolve overlapping lines on a ``fine continuum'' while doing much of the  physics on a ``coarse continuum'' \citep{2005ApJ...624..794S}. The fine continuum attempts to resolve most line profiles
while the coarse continuum has a lower resolution. The frequency-dependent line profile $k(x)$ that needs to be entered into the fine-opacity array is given by:
\begin{equation}
    k(x) = k_{\rm L} \frac{1}{\sqrt{\pi}} H(a,x) = \frac{N_1 \kappa_{\rm L}}{\Delta_{\rm v}} H(a,x),
    \label{lineprofile}
\end{equation}
where $a$ and $x$ are defined in HK80, and $H(a,x)$ is the Voigt function defined by Eq. 2.4 of the same paper. Since $H(a,0) = \exp(a^2)\,{\rm erfc}(a) \approx 1$ for $a \ll 1$, we get:
\begin{equation}
    k_{\rm L} \approx k(0) \sqrt{\pi}.
    \label{kldefalt}
\end{equation}
This latter equation can be used to generalise the treatment for overlapping lines. Instead of using Eq.~\ref{kldef3} to calculate $\beta_{HK}$, as was done in previous versions of the code\footnote{In the process of implementing this, we discovered a bug in \Cloudy{} versions C23.00 and before where the factor $\sqrt{\pi}$ was placed in the denominator rather than the numerator. This bug only affected the calculation of $\beta_{HK}$ (and hence the destruction probability), not the line profile given in Eq.~\ref{lineprofile}. This was fixed in version C23.01 \citep{2023RNAAS...7..246G}.}, we will now use Eq.~\ref{kldefalt}. This version will automatically treat line overlap when $k(0)$ is taken from the fine opacity array after all lines have been entered. One drawback of this approach is that, in general, the overlapping lines will no longer have the shape of a Voigt profile, which is implicitly assumed by HK80. 
However, since the fine-structure components are very closely spaced in \ion{H}{1} Ly$\alpha$, and in most environments the H Ly$\alpha$ lines will strongly dominate over other blended lines,
this is a minor problem for hydrogen. For high-$Z$ elements like \ion{Fe}{26} this is also a minor problem as the Ly$\alpha$ lines are clearly separated and will be treated as two distinct lines.
However, the intermediate case is more troublesome (see for instance carbon and oxygen in Figure~\ref{fig:LineOverlap}). We are not aware of a theory that presents a full treatment of line overlap in this context.

Figure~\ref{fig:LineOverlap} shows the increasing separation between the fine-structure doublets of one-electron species. 
The doublet separation is smaller than the thermal width
for small $Z$ but the lines are quite distinct at large $Z$.
The low-$Z$ elements can be treated as single lines, while the high-$Z$ and low-$n$ lines are treated as two separate lines.

The calculations of the destruction probabilities due to the continuum opacity suffer the same problem of not allowing for overlapping lines.  It is the lines from low-$Z$ ions that exhibit line overlapping. For these lines, we simply treat the doublets as two separate lines by summing the opacity of the two $j$-components to calculate both the escape and destruction probabilities. An overlapping line is considered to be one where the difference between the Doppler velocities of the two $j$-resolved lines is less than the Doppler width of the single line ($\Delta_v<\Delta$). Additional tests on the damping constant for the one-electron doublets yield that much of the line broadening that increases with Z is a result of radiative damping. 

Then using Equation~\ref{eq:detailed_balance} and the fact that $A_{4,1}$ and $A_{2,1}$ are identical to $A_{2p,1}$ (see Section~\ref{sec:transition_prob}), we calculate the unresolved escape and destruction probabilities using those for the $j$-resolved lines,
\begin{equation}
    \beta_{2p,1} = \frac{1}{3}\beta_{4,1} + \frac{2}{3}\beta_{2,1}.
\end{equation}

\subsection{Atomic Energy Levels}

Due to the high resolution of spectra that will be observed with microcalorimeter missions such as \XRISM{} and \Athena{}, a larger number of atomic high-energy levels ($n>$100) become relevant in the \Cloudy{} calculations.
\Cloudy{} by default includes only the energy levels from the Chianti v10.1 atomic database which have energies below the ionization potential of the corresponding ion (we refer to these as auto-ionizing levels further discussed in \citep{2022Astro...1..255G}). However, with spectral resolutions $R>1000$ even lines produced by transitions involving autoionizing levels will be observed. The work in \cite{2022Astro...1..255G} provides a \Cloudy{}-compatible version of the Chianti 10.0.1 database that includes all levels in the original Chianti database, available to be downloaded from \url{http://data.nublado.org/chianti/}, with filename ``chianti\_v10.1\_full''. Note, that \Cloudy{} users should use the {\tt set UTA off } command to ensure that the auto-ionizing lines are not double counted.

\subsection{The magnetic dipole line}
\label{sec:M1}

\Cloudy{} has long predicted the M1 line (the magnetic dipole $2s_{1/2} \rightarrow 1s_{1/2}$ transition), which has a transition energy and frequency very close to that of the $2p_{1/2} \rightarrow 1s_{1/2}$ transition. We update the transition energies of the M1 lines to those published in \citet{2015JPCRD..44c3103Y}.This accounts for the appropriate Lamb shift energy corrections the level and disambiguates the $2s_{1/2}$ and $2p_{1/2}$ level energies.
Furthermore, \Cloudy{} identifies specific line transitions by matching the line label and the line wavelength. The label has the typical four-character form "LLXX", where "LL" is the usual one- or two-letter element symbol, and "XX" is the ion charge, while the line wavelength has up to 6 significant figures. Thus, for transitions involving the same atom/ion with sufficiently close wavelengths, \Cloudy{} cannot disambiguate between the two lines. This was the case for the M1 line and the $2p_{1/2} \rightarrow 1s_{1/2}$ transition for low $Z$. We now disambiguate the M1 line from the fine-structure Ly$\alpha$ line within \Cloudy, by extending the line label of the former to "LLXX M1". 

The relative contributions of two-photon E1 and single-photon M1 transitions also depend strongly
on the nuclear charge, as discussed in Appendix~\ref{app:2s}.
The M1 transition is predicted to dominate for heavy elements ($Z > 40$, \citealp{1979asrt.book.....S}) such as those produced in neutron star mergers.

\section{X-ray spectral diagnostics}
\label{sec:results}

\begin{figure*}
    \centering
    \includegraphics[width=\textwidth]{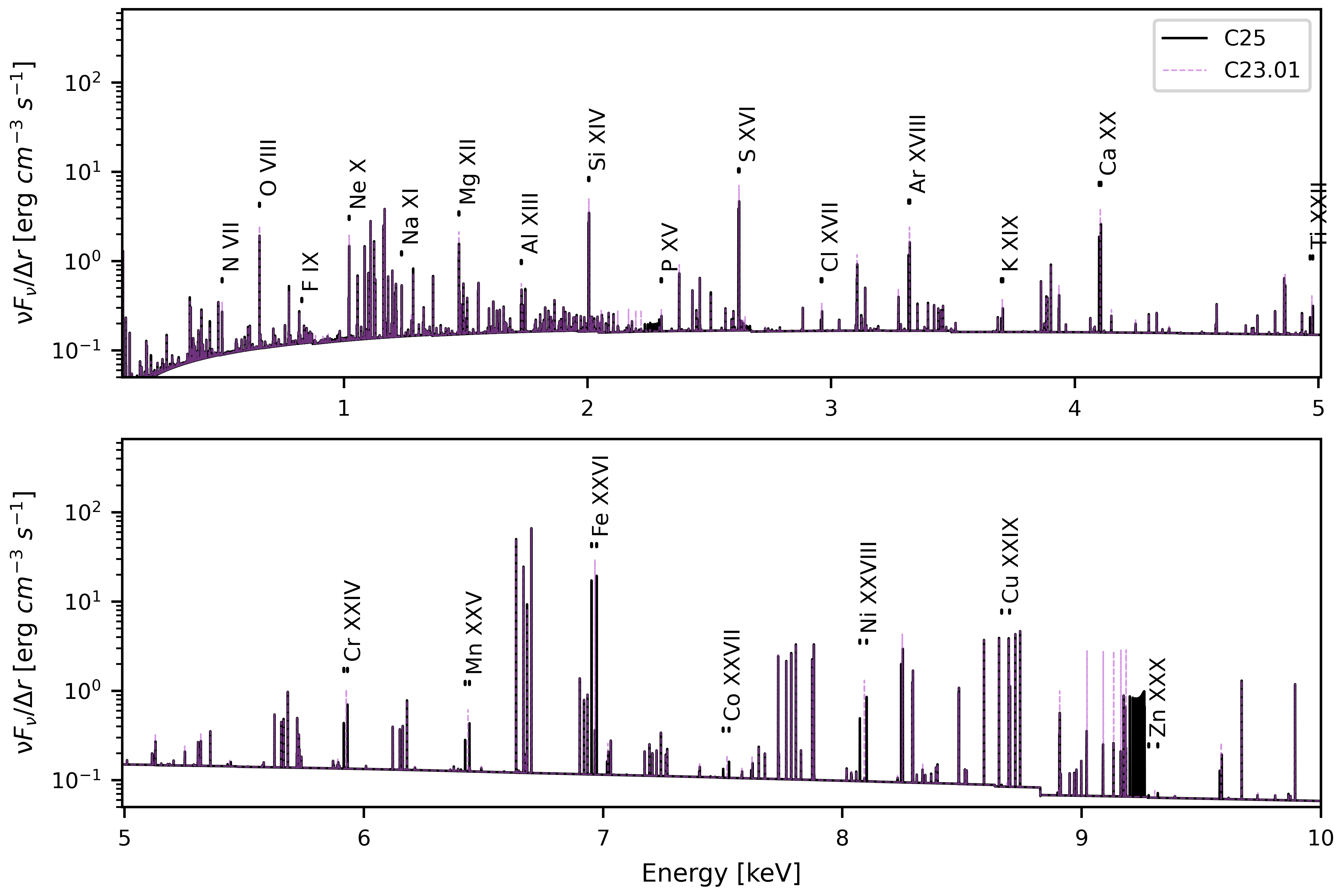}
    \caption{A \Cloudy{} simulation of the Perseus Cluster core with fine-structure doublets resolved for the one-electron ions. The black plot includes revisions that will appear in the C25 release, including the split fine-structure lines of one-electron species.}
    \label{fig:perseus}
\end{figure*}
We present \Cloudy\ calculated results, for a model of the Perseus Cluster obtained from \cite{2020ApJ...901...69C}.
Perseus being the prototypical cool-core cluster, and one of the brightest ones observed, provides an excellent model for study.
Figure~\ref{fig:perseus} shows total emission spectra, from both before and after resolving the Lyman lines. 
The model is a collisionally-ionized plasma with a constant temperature of 4.7$\times10^7$ K, hydrogen density of
$10^{-1.5}$~cm$^{-3}$, and 0.65 dex solar metallicity. A microturbulent
velocity of 150 km s$^{-1}$ is included to account for line shielding and pumping.
The energy range, 0.4 - 10 keV, is covered by the \XRISM{} mission, and includes the important Fe Ly$\alpha$ doublet. The y-axis has been scaled by the thickness of the cloud $\Delta r$, since the line intensities will depend linearly on the $\Delta r$ for optically thin transitions with no external radiation field. 

With the default spectral resolution implemented (1/10 of \Athena's resolution, 0.25 eV) we see that \Cloudy{} now predicts Ly$\alpha$ doublets for all H-like species heavier than nitrogen. Since this update provides an energy accuracy beyond the current and future X-ray microcalorimeter spectral resolution (as shown by Figure~\ref{fig:dE}), the improved \Cloudy{} spectra will be in excellent agreement with microcalorimeter observations.

The following subsections use the Perseus cluster core model presented above. We expand the model for the range of hydrogen column densities that occur in X-ray emitting clouds, $18<\log$ $N$(H)$<25$. The input scripts and the ensuing figures are available in \href{https://gitlab.nublado.org/cloudy/papers}{gitlab.nublado.org/cloudy/papers}. The choice of including \ion{Ca}{20} 
in the following subsections is justified by the fact that \XRISM/Resolve will resolve Ly$\alpha$ doublets only for elements at least as heavy as calcium \citep{2025A&A...694L..13G}. 

\subsection{GHF Diagnotic Diagrams}
\begin{figure*}
    \centering
    \begin{subfigure}[t]{0.48\textwidth}
        \centering
        \includegraphics[width=\textwidth]{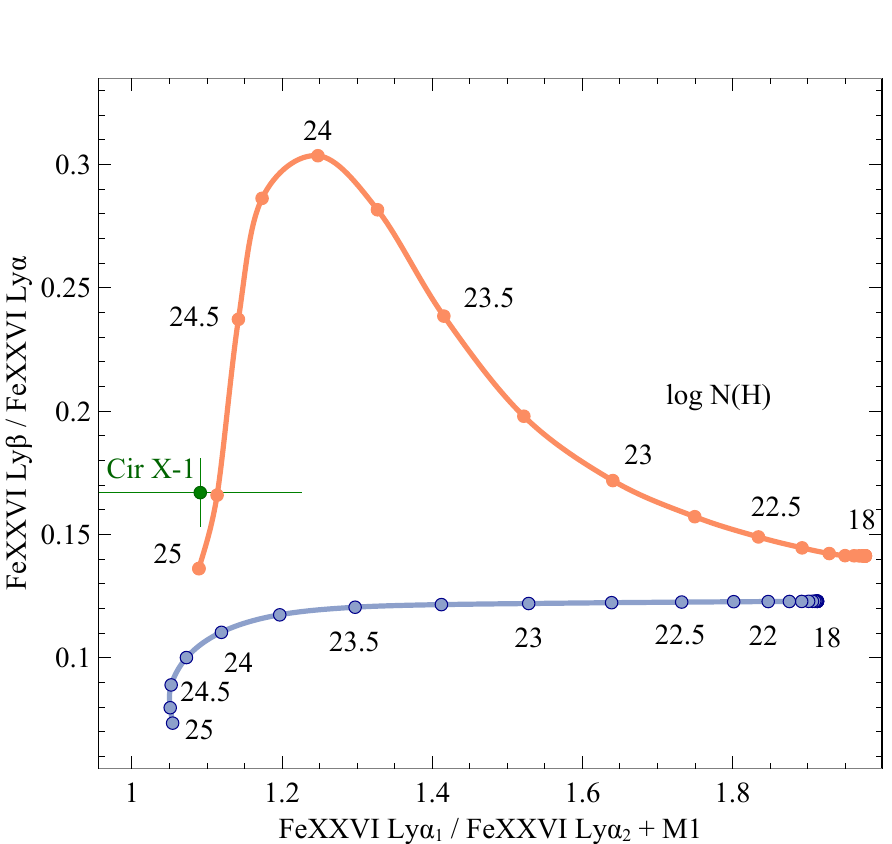}
        \caption{}
        \label{fig:BPT:panel1}
    \end{subfigure}
    \hfill
    \begin{subfigure}[t]{0.48\textwidth}
        \centering
        \includegraphics[width=\textwidth]{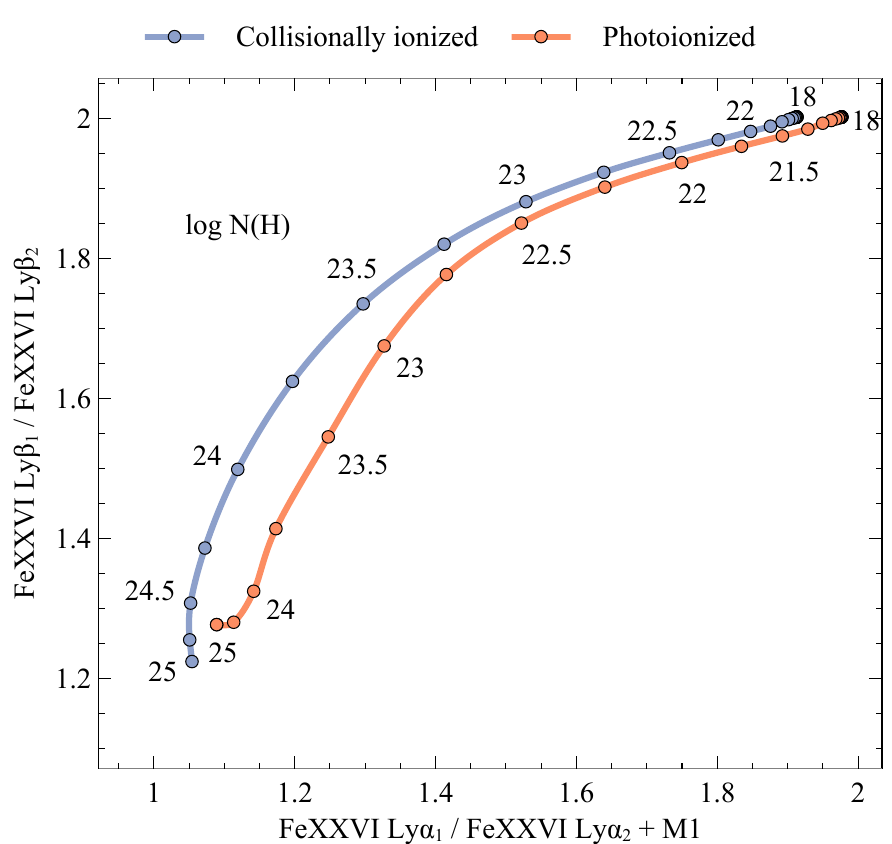}
        \caption{}
        \label{fig:BPT:panel2}
    \end{subfigure}
    \caption{{A GHF diagnostic diagram (a BPT-like diagram for X-ray emission lines) using \Cloudy{} models of the Perseus Cluster core which is collisionally ionized and model of a high-mass X-ray binary which is photoionized. Here we show the \textit{(a)}: Lyman decrement and \textit{(b)}: the j-resolved Ly$\beta$ ratio against the j-resolved doublet ratio of Ly$\alpha$ as hydrogen column density is varied. For illustrative purposes, panel \textit{(a)} shows in green an XRISM/\textit{Resolve} observation of Circinus X-1, obtained from \citet{2025PASJ..tmp...30T}. These are robust column density diagnostics.}}
    \label{fig:BPT}
\end{figure*}

{BPT diagrams are a fundamental diagnostic tool used in UV and optical regimes, to distinguish astrophysical objects based on their emission line ratios \citep{1981PASP...93....5B, 1987ApJS...63..295V}. 
XRISM's spectral resolution provides X-ray astronomy with an opportunity to adopt similar diagnostic techniques. 
We present here a novel BPT-like diagnostic tool to be utilized for probing X-ray sources, hereby referred to as GHF (Gunasekera, van Hoof, and Ferland) diagnostic diagrams.} 

{Figure\,\ref{fig:BPT} shows two examples of such a GHF diagram, expanding on our previous work in \citep{2025A&A...694L..13G}. 
The figure presents the model of a high-mass X-ray binary as the photoionized model obtained from \citet{2025A&A...694L..13G} and the above described Perseus cluster core model as the collisionally ionized one. 
We show the change in the intensity ratio of the \ion{Fe}{26} Lyman decrement (Ly$\beta$/Ly$\alpha$) (Figure\,\ref{fig:BPT:panel1}) and the \ion{Fe}{26} Ly$\beta$ doublet (Figure\,\ref{fig:BPT:panel2}) against that of the \ion{Fe}{26} Ly$\alpha$ doublet, as we vary the hydrogen column density. The green point in the figure represents an outflowing photoionized plasma from Circinus X-1, obtained from \cite{2025PASJ..tmp...30T}, observed by XRISM/\textit{Resolve}. 
}

{Both the Ly$\alpha$ and Ly$\beta$ lines are useful probes of the physical conditions of the X-ray emitting plasma, and together they provide a robust method to constrain the hydrogen column density. The proceding subsections explain the physics behind the Case A to Case B variation in the Ly$\alpha$ and Ly$\beta$ ratios. 
Several observations using XRISM/Resolve have already observed a Ly$\alpha_1$/Ly$\alpha_2$ ratio which deviates from the expected 2:1 ratio \citep{2025PASJ..tmp...30T, 2025ApJ...985L..20X, 2025ApJ...982L...5X, 2025arXiv250506533X}. These deviations are a result of a change to the optical depths of the $j$-resolved components of Ly$\alpha$, which traces the H column density of the corresponding gas.
}

{The GHF plots show that this column density diagnostic can be used for both a collisionally ionized or a photoionized gas. 
Assuming that gas parameters and the ionizing radiation of the outflowing plasma from Circinus X-1 match the model parameters presented in Figure\,~\ref{fig:BPT:panel1}, the plot tells us that this plasma should have a ionization parameter close to $10^5$ erg\,\(\mathrm{cm}\,\mathrm{s}^{-1}\) and H column density ~$5\times10^{24}$ \(\mathrm{cm}^{-3}\). We use Cir X-1 observation here to illustrate the diagnostic power of these plots.
The script for these models are available in \url{https://gitlab.nublado.org/cloudy/papers}, which can be obtained to conduct a broader analysis of Cir X-1.
Other gas parameters that affect the optical depth of X-ray emission lines, such as the ionization parameter, will also make excellent candidates for GHF diagnostic diagrams. 
}

{A major difference between the GHF plots in panels (a) and (b) of Figure\,\ref{fig:BPT} is that, in the former the photoionization curve diverges significantly from the collisionally ionized curve. So an observational point placed on (a) will vary much more drastically based on the ionization mechanism, for a gas with the same $N$(H), than it would in Figure (b). So an observation could identify the cloud's energy source. Future work exploring these parameter spaces is underway to get a deeper understanding of them. One example is whether panel (b) is a more appropriate diagnostic tool than (a) in cases where the ionization mechanism is unknown.
}

\subsection{A Column Density Indicator}

\begin{figure}
    \centering
    \includegraphics[width=\columnwidth]{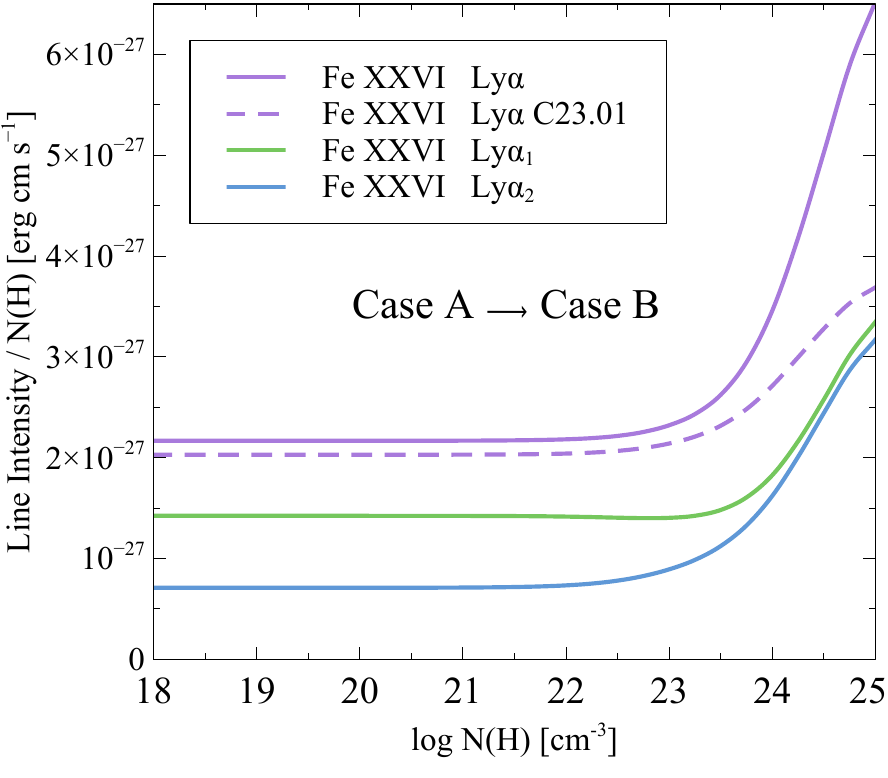}
    \caption{This shows the intensity of the
    \ion{Fe}{26} Ly$\alpha_1$ and
    Ly$\alpha_2$ lines, as well as the new unresolved Ly$\alpha$, alongside the unresolved Ly$\alpha$ as predicted using C23.01. We show the individual line intensities divided by the gas column density as a function of column density.}
    \label{fig:Ka1_Ka2_intensity}
\end{figure}

We present the classical \citet{1938ApJ....88...52B} Case A (small $N$) to Case B (large $N$) limits \citep{2006agna.book.....O, 2021ApJ...912...26C}, for the Perseus cluster core model.
This is a collisionally ionized plasma, so photoionization and radiative
excitation are not important.
Figure~\ref{fig:Ka1_Ka2_intensity} shows the total Fe Ly$\alpha$, the j-resolved Ly$\alpha_{1,2}$ doublet, along with their counterpart that is predicted by the previous version of \Cloudy{}, C23.01. 
The figure shows sensitivity of the Fe Ly$\alpha$ l to H column densities $>10^{22}$ cm$^{-3}$. 
The results with the j-resolved doublet reveal that Ly$\alpha$ is now much stronger compared to the C23.01 model. There was a clear underestimation of the intensity of the \ion{Fe}{26} Ly$\alpha$ line predicted in C23.01.
The total Ly$\alpha$ optical depth is larger in the
older unresolved model since the total opacity of the line is the sum of the opacities of the two fine-structure components.
With the $j$-resolved model, the two components of Fe {\sc xxvi} Ly$\alpha$ do not overlap (see Figure\,\ref{fig:LineOverlap}) so the optical depth and opacity are smaller and photons escape more freely.
As we will show in Section~\ref{sec:CaseAtoB}, the Case A to B transition enhances the Ly$\alpha$ line at large column densities.
As $N$(H) increases, so do the line optical depths. Scattering Ly$\beta$ photons have a finite probability of being absorbed and re-emitted as Ba$\alpha$ followed by Ly$\alpha$, intensifying both the Ly$\alpha$ and Ba$\alpha$ lines \citep{1985ApJ...299..752N, 2015tsaa.book.....H}.
The weaker Ly$\alpha$ line in C23.01 implies that Ly$\beta$ destruction dominates over Ly$\alpha$ destruction much more than previously calculated.

\subsection{Transitions from Case A to Case B}
\label{sec:CaseAtoB}

\begin{figure}
    \centering
    \includegraphics[width=\columnwidth]{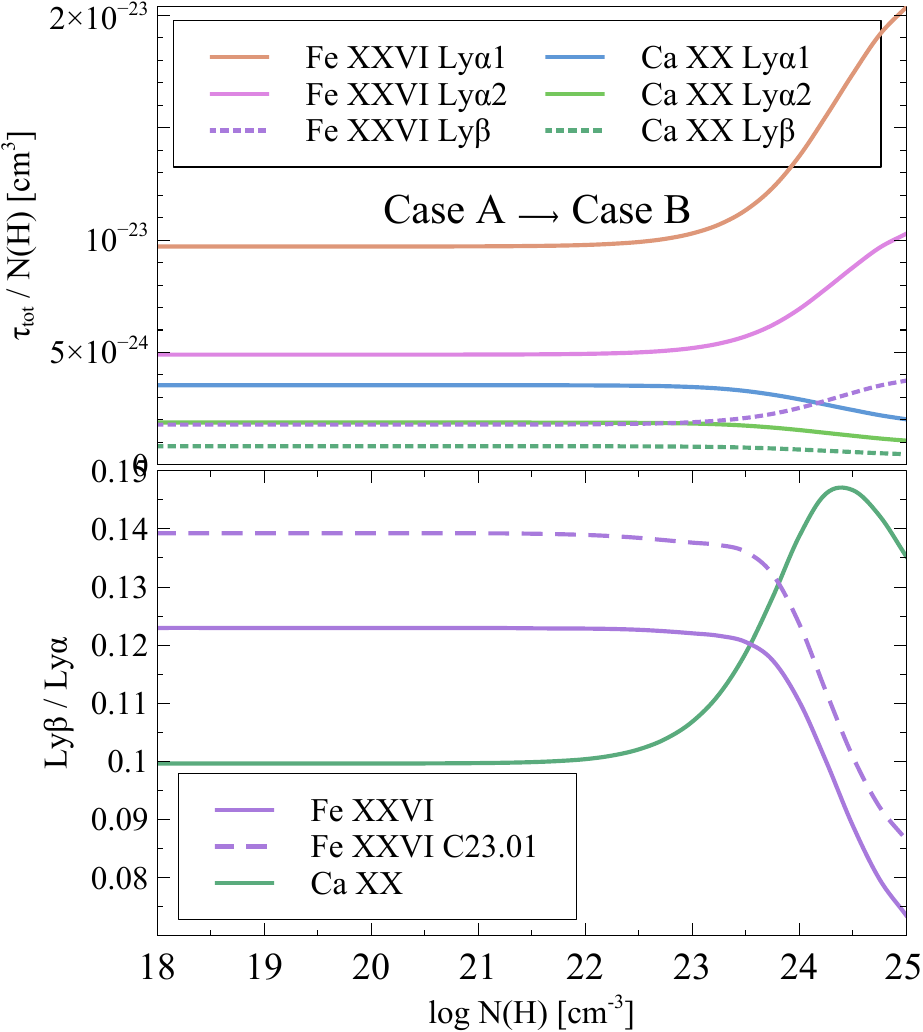}
    \caption{\Cloudy{} models of the Perseus Cluster core. \textit{Top panel:} Line optical depth per unit column density, as a function of column density. 
    \textit{Bottom panel:} The intensity ratio of the
    \ion{Fe}{26} and \ion{Ca}{20} Ly$\beta$ to
    Ly$\alpha$ lines. This is a column
    density diagnostic because significant
    line optical depths cause the Ly$\beta$ to
    undergo multiple scattering and be
    transformed to the Ba$\alpha$
    and Ly$\alpha$ lines. This is the classical \citet{1938ApJ....88...52B} Case A to B limit. The details of the behaviour in this figure are model dependent and are not universal.}
    \label{fig:FeKbovKa}
\end{figure}

The higher-$n$ Lyman lines also provide a column density indicator. Figure\,\ref{fig:FeKbovKa} shows the predicted Ly$\beta$/Ly$\alpha$ intensity ratio as a function of the column density (bottom panel), along with optical depth per unit column density of each line (top panel). We report both Ly$\beta$ ($n=3 \rightarrow 1$) and Ly$\alpha$ ($n=2 \rightarrow 1$) as the multiplet sum in this Figure.

The dependencies of calcium and iron ionization are complex.
The $\tau$/$N$(H) of an emission line changes with total column density, as shown in the top panel of Figure~\ref{fig:FeKbovKa}. 
This reveals that the mean ionization changes as $N$(H) increases. Large column densities have large optical depths, resulting in trapped line radiation within the cloud. 
These trapped lines, scatter many times photoionizing other elements, affecting the mean ionization of the cloud. Here, the ionization of the H-like Fe increases with $N$(H), while that of H-like Ca decreases, resulting in the difference in the Ly$\beta/$Ly$\alpha$ trends between \ion{Fe}{26}, and \ion{Ca}{20} with $N$(H). 

The Case A to B transition increases the optical depth in the lines. Each Fe Ly$\beta$ photon scattering has a $\sim 11$\% probability (this probability comes from the Einstein $A_{u,l}$ ratio $\frac{A_{3p,2s}}{A_{3p,1s} + A_{3p,2s}}$) of being converted into Ba$\alpha$ ($n=3\rightarrow2$) followed by Ly$\alpha$. Multiple scattering causes the \ion{Fe}{26} Ly$\alpha$ to grow stronger and Ly$\beta$ weaker. This Fe Ly$\alpha$ to Fe Ly$\beta$ conversion begins as Ly$\beta$ weakens at $N($H$)\sim 10^{21}$~cm$^{-3}$ and provides a column density diagnostic. However, the Ca Ly$\beta/$Ly$\alpha$ ratio, for this particular model, exhibiting a maximum as a function of $N$(H) is not ideal as a column density indicator.

\subsection{Case C to Case B}
Case C is the limit where the Ly$\alpha$ lines are optically thin and the incident radiation field can pump the lines \citep{1938ApJ....88..422B}. This fluorescent excitation makes the lines stronger \citep{1999PASP..111.1524F}. It occurs in photoionized clouds with lower column densities. As the cloud column density increases, the Ly$\alpha$ line optical depth increases and the transition becomes self-shielded from the incident radiation field. The emission goes over to the Case B limit \citep{1938ApJ....88...52B, 2021ApJ...912...26C}.

\begin{figure}
    \centering
    \includegraphics[width=\columnwidth]{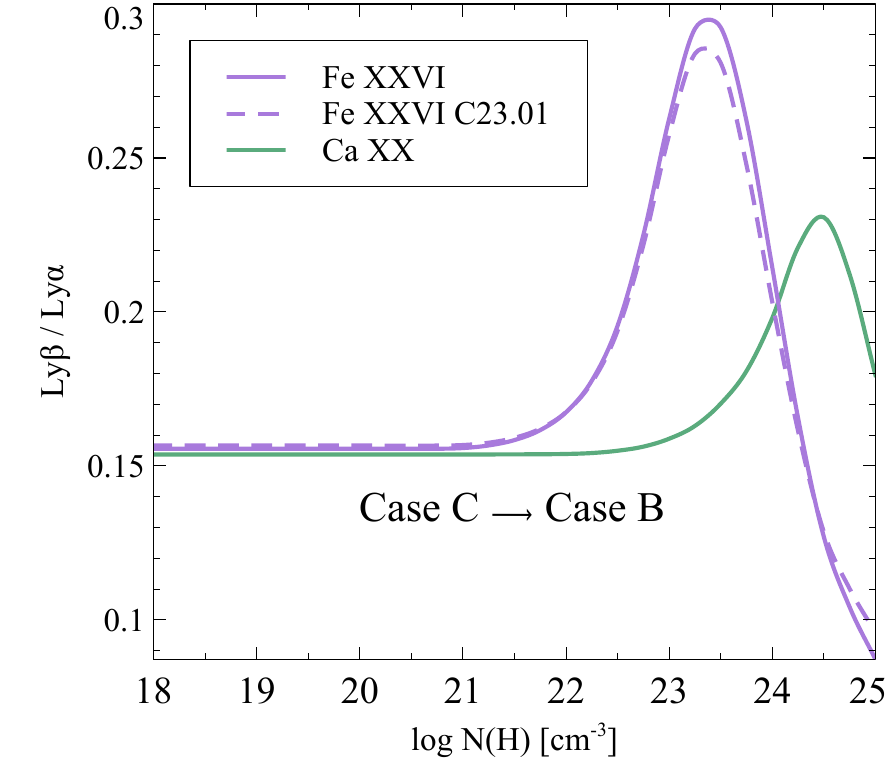}
    \caption{\Cloudy{} models for the Perseus Cluster core showing the intensity ratio of the
    \ion{Fe}{26} and \ion{Ca}{20} Ly$\beta$ to
    Ly$\alpha$ lines. Just as Figure
    \ref{fig:FeKbovKa} shows the
    Case A to Case B transition for a collisionally
    ionized gas, this shows the Case C 
    \citep{1938ApJ....88..422B} to Case B transition for gas illuminated by an SED that fluoresces the one-electron Lyman series.
    This is also a column
    density diagnostic. }
    \label{fig:CaseC}
\end{figure}

Figure\,\ref{fig:CaseC} shows a series of calculations based on those shown in Figures \ref{fig:Ka1_Ka2_intensity} and 
\ref{fig:FeKbovKa}. {It assumes a constant gas kinetic temperature of $4.7\times 10^7$~K for simplicity but is also exposed to a powerlaw SED with an ionization parameter of $\log U = 3$.} This radiation field is strong enough to pump the Ly$\alpha$ lines but not so strong as to change the ionization of the cloud. This is done to expose the essential physics and is not meant as a realistic model of a photoionized cloud. In a true photoionized cloud, the kinetic temperatures and ionization would change as a function of column density, obfuscating the essential physics.

\section{Conclusions}

With the launch of the X-ray observatory \XRISM\ with spectral resolution $R>1000$, 
the need arises for analysis tools such as \Cloudy{} to make full use of these new data. 
The series of thesis papers by Priyanka Chakraborty
\citep{2020ApJ...901...68C, 2020ApJ...901...69C, 2021ApJ...912...26C, 2022ApJ...935...70C}
expanded our treatment of two-electron spectra such as \ion{Fe}{25}. 
This project has advanced \Cloudy{}'s one-electron iso-sequence spectral-line predictions to 
match that of microcalorimeter observations, in a manner that is self-consistent 
with the microphysics in the cloud. 
This paper described how one-electron spectra are predicted in \Cloudy{}.

We used atomic theory rather than database lookup for the
atomic framework of our model atoms. 
This ensures completeness and extensibility.
\Cloudy\ treats species along one- and two-electron isoelectronic
sequences with a unified model.
We currently include all elements between hydrogen and zinc.
Our theoretical approach for the atomic data ensures that we have
complete coverage of all of these elements.
The treatment is general, so it will be straightforward to
extend our predictions to very heavy elements such as those produced in 
neutron star mergers.

We extended the line redistribution theory used in \Cloudy{} for overlapping lines.
Our framework was originally developed to treat
the strongly overlapping H$_2$ electronic
lines using a multi-grid approach
\citep{2005ApJ...624..794S}.
Figure \ref{fig:energyResPlot} shows that
the degree of overlap of the one-electron doublet depends on the nuclear charge and the local velocity field. 
In the course of this development, we discovered a simple coding error which
was described and corrected in \citet{2023RNAAS...7..246G}.
Our treatment remains general and can be applied to any set of overlapping lines.

We identify the lack of data for $j$-changing collisions within the $n=2$ shell 
as the most pressing missing part of our simulations. 
High-quality electron collision data now exist \citep{2022ApJS..263...35M}. Slow-moving particles such as protons or alpha particles have the largest cross sections for $j$- and $l$-changing collisions. These exist only for a few ions. 
This is discussed in Section~\ref{sec:pop_den}). To include the one-electron $j$-resolved $2p_{1/2,3/2} \rightarrow 1s_{1/2}$ transitions in \Cloudy{}'s full CRM solver, and provide more accurate spectral predictions for the next generation of X-ray observations, it is essential to fill this data gap.

Lastly, Fe Ly$\alpha$ has long been an important measurement in X-ray observations. Fe being abundant in high-energy environments, the Ly$\alpha$ line can provide valuable information on the physical conditions of such regions. Using the work from this study, we show in \cite{2025A&A...694L..13G} that the Fe Ly$\alpha_1/\alpha_2$ can be a powerful column density indicator. This work will highlight some insightful physics we can probe as a result of being able to resolve Ly$\alpha$ into the doublet, with \XRISM observations. 

The scripts for all figures presented in this paper are found in the directory named after the bibcode of the present paper, in our openly accessible repository \href{https://gitlab.nublado.org/cloudy/papers}{gitlab.nublado.org/cloudy/papers}.
These make it easy for others to build on our modelling.

\begin{acknowledgments}
We would like to acknowledge Stefano Bianchi for his advice, and Masahiro Tsujimoto for his insight on XRISM observations, and all the participants at the \Cloudy{} 2024 workshop in Tokyo, for testing this development of the code.
CMG and GF acknowledges support from JWST-AR-0628, JWST-AR-06419, and NSF (1910687).
MC acknowledges support from NASA (19-ATP19-0188, 22-ADAP22-0139) and NSF (1910687).
\end{acknowledgments}

\begin{contribution}

All authors contributed equally.

\end{contribution}

%



\appendix

\section{Energy Calculations}
\label{app:energy}

The following energy level corrections are the complete calculations performed by \Cloudy{} on the terms introduced in Eq.~\ref{eq:totalE}.

\subsection{Fine-Structure Splitting}

Consider a single electron orbiting an atomic nucleus of charge $Z$. An accelerating charge sets up a magnetic field. This field exerts a torque on the magnetic moment of the nucleus, causing it to align with the field. Hence the general Hamiltonian of the electron in the magnetic field of the nucleus is,
\begin{equation}
    H = \mathbf{\mu}_e.\mathbf{B}_N.
\end{equation}

The magnetic field of the nucleus can be written in relation to the orbital angular momentum of the electron. In the rest frame of the electron, the magnetic field can be approximated by a current loop. The current is given by $I=Ze/T$, where $T$ is the orbit period. Since this is the same orbital period of the electron which relates to its orbital angular momentum, $\mathbf{B}_N \propto \mathbf{L}_e$.

The magnetic dipole moment of a spinning charge is related to its spin angular momentum. So, we have $\mathbf{\mu}_e = -\frac{e}{m} \mathbf{S}_e$. Hence $H\propto\mathbf{S.L}$, which is the spin-orbit interaction. The eigenvalues of this term are given by $j(j+1)-l(l+1)+s(s+1)$. For an electron, we have $s=1/2$, so there is no dependence on $s$. Additionally, the relativistic correction cancels out the orbital angular momentum quantum numbers, leaving only a dependence on the quantum number $j$. The level energy including all these corrections with the fine structure included are,
\begin{equation}
    E_n^0 + E^{FS}_{nj} = m_e c^2\left[1+ \left( \frac{\alpha Z}{n-k+\sqrt{k^2 - \alpha^2 Z^2}} \right)^2 \right]^{-\frac{1}{2}} - m_e c^2 
    \label{eq:dirac}
\end{equation}
where, $j\in\{|1/2-l|,...,(1/2+l)\}$, $k=j+1/2$, $n$ is the principal quantum number, $\alpha$ is the fine structure constant and $m_e$ is the mass of the electron. $E^0_n$ is the unperturbed energy.
This interaction can be thought of as a perturbation that partially lifts the degeneracy of the energy states by splitting the levels with different orbital quantum number $l$. For all one-electron systems, the $^2$P state ($l=1$) is resolved into a doublet with $j=1/2, 3/2$.

\subsection{Other \texorpdfstring{$np$}{np} Level Energy Corrections}
The Lamb Shift correction, $E_{n,l>0,j}^{\rm LS}$, resolves the levels with different $l$,
\begin{equation}
    E_{n,l>0,j}^{\rm LS} = \frac{8Z^4\alpha^3}{3\pi n^3}Ry\left[\log{\frac{Z^2Ry}{K_0(n,l)}}+\frac{3}{8}\frac{c_{lj}}{2l+1}\right],
    \label{eq:E_LS}
\end{equation}
\begin{equation}
c_{lj}=
    \begin{cases} 
(l+1)^{-1}, & \mbox{j=l+1/2}, \\ 
-l^{-1}, & \mbox{j=l-1/2}.
\end{cases}
\end{equation}
where $\log{K_0(n,l=1)/Z^2Ry}$ is the Bethe logarithm. The numerical value of $K_0(n,l=1)$ is difficult to evaluate for a large number of $n$, so we developed an approximation of the Bethe logarithm as discussed in Appendix~\ref{sec:Bethe_log}. For the present purpose, this sufficiently satisfies the current and future instrumental needs as discussed in the following section.

The nuclear mass recoil correction $E_{nj}^{M}$, is
\begin{equation}
    E_{nj}^{M} = m_ec^2\frac{m_e}{m_N}\frac{(\alpha Z)^2}{2N^2}-\mu c^2\left(\frac{m_e}{m_N}\right)\frac{(\alpha Z)^2}{2n^2}
    \label{eq:E_M}
\end{equation}
\begin{equation}
    N = \left(\left(n-k+\sqrt{k^2-\alpha^2Z^2}\right)^2+\alpha^2Z^2\right)^{1/2}
    \label{eq:N}
\end{equation}
where $k=j+1/2$ as before \citep{2015JPCRD..44c3103Y}. Known also as the mass shift correction, this corrections accounts for the slight recoil of the nucleus from attraction to the electron.

\subsection{Bethe Logarithm}
\label{sec:Bethe_log}
The Bethe logarithm $\ln(K_0/Z^2Ry)$ is a dimensionless quantity, where $K_0(n,l)$ represents the mean excitation energy for the Lamb Shift. It was first introduced by Hans Bethe in 1947 as part of his theory of the Lamb shift \citep{1957qmot.book.....B}. For one-electron atoms with $l\ne0$, the Bethe logarithm requires the evaluation of oscillator strengths for transitions $nl \rightarrow n',l\pm1$. Such a calculation can become tedious when evaluating the Bethe logarithm for a large value of $n$. Since $K_0(n,l\ne0)/Z^2Ry$ varies slowly with $n$, and does not reduce to values much smaller than unity for $n\rightarrow\infty$, we can approximate it with a negative exponential. 
\begin{figure}
    \centering
    \includegraphics[width=0.6\columnwidth]{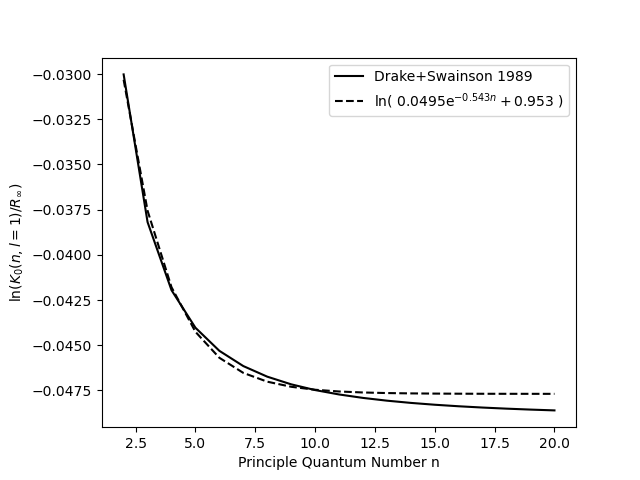}
    \caption{Derived best fit line to the $K_0(n,l)$ term of the Bethe logarithm.
    }
    \label{fig:BetheLog}
\end{figure}
Using an optimized curve fit to the numerical evaluations of the Bethe logarithm presented in \citet{1990PhRvA..41.1243D}, for $l=1$, we find the following approximation,
\begin{equation}
    \frac{K_0(n,l=1)}{Z^2Ry} \approx 0.0495e^{-0.543n} + 0.953.
    \label{eq:Bethe_log}
\end{equation}
Figure~\ref{fig:BetheLog} shows our fit evaluated for $n$ up to $n=20$ and compared with the values in \citet{1990PhRvA..41.1243D}. We find that Equation~\ref{eq:Bethe_log} provides a sufficiently good fit, especially for low $n<10$, with a mean square error of order $10^{-7}$.

\section{The one-electron forbidden transitions}
\label{app:2s}

\Cloudy{} considers all elements up to $Z=30$ (zinc). 
The structure of the code is designed to be readily expandable to heavier
elements, so development could extend to $Z>30$. 
The initial emphasis was on ultraviolet and optical spectroscopy,
as driven by the instrumentation available at the time.
The new generation of X-ray microcalorimeters has
motivated our recent development of the one and two-electron
systems at high $Z$. 
Previous papers focused on two-electon systems, with an 
emphasis on iron 
\citep{2020ApJ...901...68C, 2020ApJ...901...69C, 2021ApJ...912...26C, 2022ApJ...935...70C}.

Here we point out an interesting aspect of the $2s_{1/2} \rightarrow 1s_{1/2}$
transition. There are no allowed (E1) single-photon transitions between these two states,
but so-called ``forbidden'' transitions are possible.
For galactic nebulae, this produces the strong \ion{H}{1} two-photon continuum
\citep{2006agna.book.....O}.
Various moments of the radiative transition rates have
different dependencies on charge \citep{2020ApJ...901...69C}.
Here, we show how the types of emission produced by the
$2s_{1/2} \rightarrow 1s_{1/2}$ M1 transition change with Z.
At low $Z$ the two-photon (2E1) continuum is dominant \citep{2006agna.book.....O}, 
while at very high $Z$ the single magnetic dipole (M1) line dominates.

Single-photon electric dipole transitions would violate the parity selection rule, but two-photon
electric dipole transitions can occur.
This produces a broad continuum that, in terms of photon number, peaks
at half the energy of Ly$\alpha$, 

Section 11.2.1 of \citet{1979asrt.book.....S} discusses the
$2s_{1/2} \rightarrow 1s_{1/2}$ transition in detail \citep{1981PhRvA..24..183G}.
\begin{equation}
    2E1:A(2s \rightarrow 1s) =
    8.230 \ Z^6\,
    \frac{1+3.95(\alpha Z)^2-2.04(aZ)^4}
    {1+4.60(aZ)^2} \
    [s^{-1}]
\end{equation}

Magnetic dipole transitions are also possible between the two states, but are
much weaker than the two-photon transitions for low-charge elements.
However, the magnetic dipole transitions have a much stronger dependence on $Z$ \citep{1979AdAMP..14..181M}:
\begin{equation}
M1:A(2s \rightarrow 1s) =
\frac{\alpha^9 Z^{10} }{ 972}
\frac{m e^4}{h^3}
\approx 2.46\times 10^{-6} Z^{10} \
[s^{-1}]
\end{equation}

\begin{figure}
    \centering
    \includegraphics[width=0.6\columnwidth]{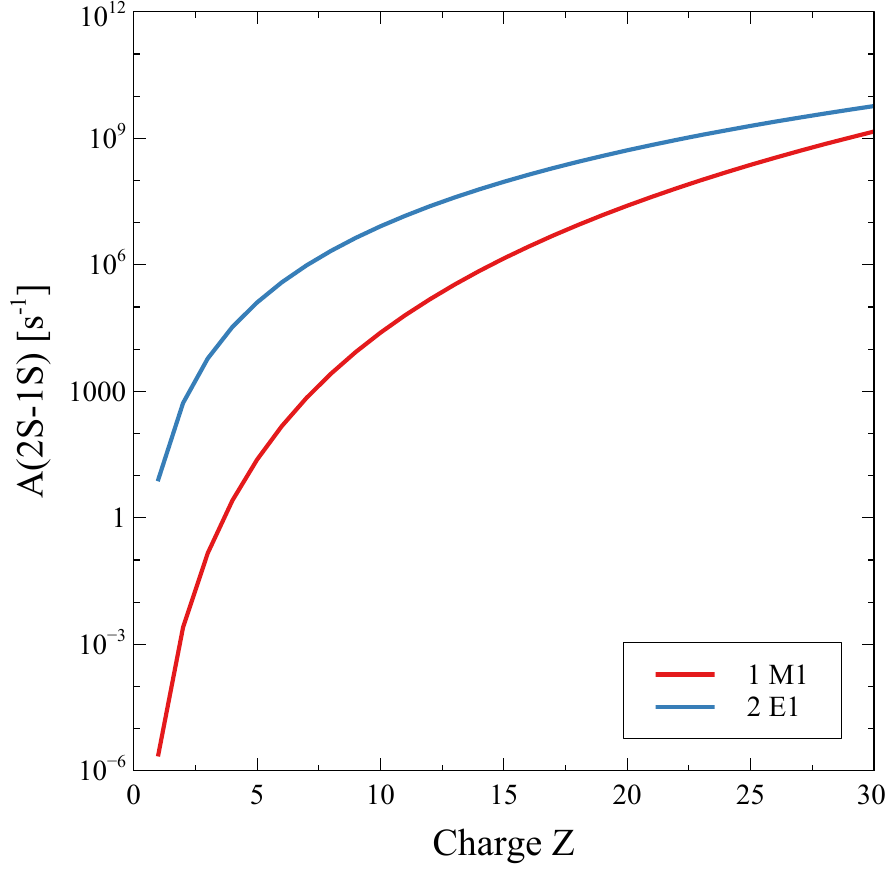}
    \caption{Change in the transition rate due to one and two 
    photon processes. At low-$Z$, two-photon emission is $\sim$7 dex
    faster than the $M1$ transition while the two are
    comparable for Fe-peak elements.}
    \label{fig:SobelmanE1M1}
\end{figure}

We report both the one- and two-photon emission for all $Z$.
Two-photon emission dominates at low $Z$, and we report that with the 
species name (i.e., ``H~~1'') but with a wavelength twice
that of Ly$\alpha$ (since that is where the two-photon continuum peaks).
For iron-group elements, the one and two-photon rates are comparable but do
not present a spectroscopic blend since they present line and 
continuum emission.
We report this with the species name with M1 appended (i.e., ``H~~1~M1'') and the wavelength of the M1 transition, which is very close to the $2p_{1/2} \rightarrow 1s_{1/2}$ transition.
We also report a blend (with species label "Blnd") of sum of the the M1 and $2p_{1/2} \rightarrow 1s_{1/2}$ transitions, and another of the M1, and both $2p \rightarrow 1s$ transitions.


\bibliography{Xray-Cloudy}{}
\bibliographystyle{aasjournalv7}



\end{document}